\documentclass[twoside,twocolumn]{article}

\usepackage[super,sort&compress,comma]{natbib}
\usepackage{amsmath}
\usepackage{amssymb}
\usepackage{booktabs}
\usepackage{times,mathptmx}
\usepackage{sectsty}
\usepackage{balance} 
\usepackage{graphicx,color}         
\usepackage{mathrsfs}
\usepackage{lastpage}
\usepackage[format=plain,justification=raggedright,singlelinecheck=false,font=small,labelfont=bf,labelsep=space]{caption} 
\usepackage{fancyhdr}
\usepackage[normalem]{ulem}

\begin{document}


\oddsidemargin -1.2cm
\evensidemargin -1.2cm
\textwidth 18cm
\headheight 1.0in
\topmargin -3.5cm
\textheight 22cm

\widowpenalty=100
\clubpenalty=100

    \renewcommand{\topfraction}{0.8}	
    \renewcommand{\bottomfraction}{0.7}	
    \setcounter{topnumber}{2}
    \setcounter{bottomnumber}{2}
    \setcounter{totalnumber}{4}     
    \setcounter{dbltopnumber}{2}    
    \renewcommand{\dbltopfraction}{0.9}	
    \renewcommand{\textfraction}{0.07}	

\thispagestyle{plain}
\renewcommand{\thefootnote}{\fnsymbol{footnote}}
\renewcommand\footnoterule{\vspace*{1pt}%
\hrule width 3.4in height 0.4pt \vspace*{5pt}} 
\setcounter{secnumdepth}{5}

\makeatletter 
\def\subsubsection{\@startsection{subsubsection}{3}{10pt}{-1.25ex plus -1ex minus -.1ex}{0ex plus 0ex}{\normalsize\bf}} 
\def\paragraph{\@startsection{paragraph}{4}{10pt}{-1.25ex plus -1ex minus -.1ex}{0ex plus 0ex}{\normalsize\textit}} 
\renewcommand\@biblabel[1]{#1}            
\renewcommand\@makefntext[1]%
{\noindent\makebox[0pt][r]{\@thefnmark\,}#1}
\makeatother 
\renewcommand{\figurename}{\small{Fig.}~}
\sectionfont{\large}
\subsectionfont{\normalsize} 

\renewcommand{\headrulewidth}{1pt} 
\renewcommand{\footrulewidth}{1pt}
\setlength{\arrayrulewidth}{1pt}
\setlength{\columnsep}{6.5mm}
\setlength\bibsep{1pt}

\newcommand{\ml}[1]{\MakeLowercase{#1}}


\newcommand{\bvec}[1]{\ensuremath{\mathbf{#1}}}

\newcommand{\fig}[1]{Fig.~\ref{#1}}
\newcommand{\eqn}[1]{Eqn.~(\ref{#1})}
\newcommand{\tableref}[1]{Table~\ref{#1}}

\newcommand{\tab}{\hspace{3ex}}
\newcommand{\noi}{\noindent}

\setlength{\tabcolsep}{1.5 ex} 
\newcommand{\mctwo}[1]{\multicolumn{2}{c}{#1}}
\newcommand{\mcc}{\mctwo{$c$}}

\newlength{\figwidth}
\setlength{\figwidth}{3.3in}

\newenvironment
	{revtabular}[1]
	{\begin{tabular}{#1} \toprule }
	{\\ \bottomrule \end{tabular}}

\newcommand{\nb}[1]{\textcolor{blue}{#1}}



\newcommand{\bra}[1]{\left \langle #1 \right |}
\newcommand{\ket}[1]{\left | #1 \right \rangle}

\newcommand{\dr}{d^3\mathbf{r}}

\renewcommand{\k}{\ensuremath{\kappa}}		
\newcommand{\K}{\ensuremath{K}}				
\newcommand{\D}{\ensuremath{D}}				
\newcommand{\z}{\ensuremath{\zeta}}			

\newcommand{\Nnorm}{\ensuremath{N}}
\newcommand{\EN}{\ensuremath{E_{N}}}

\newcommand{\T}{\ensuremath{T}}

\newcommand{\ldb}{\lambda_\mathrm{dB}}

\newcommand{\mathhe}{\mathrm{He}}
\newcommand{\xhe}{\ensuremath{X\mathrm{He}}}
\newcommand{\nhe}{\ensuremath{n_\mathrm{He}}}
\newcommand{\nxhe}{\ensuremath{n_{\xhe}}}
\newcommand{\nx}{\ensuremath{n_X}}
\newcommand{\tf}{\ensuremath{\tau_\mathrm{f}}}
\newcommand{\td}{\ensuremath{\tau_\mathrm{d}}}
\newcommand{\tc}{\ensuremath{\tau_\mathrm{c}}}
\newcommand{\ag}{\textrm{Ag}}
\newcommand{\he}{\textrm{He}}
\newcommand{\hethree}{\ensuremath{{^3}}\textrm{He}}
\newcommand{\hefour}{\ensuremath{{^4}}\textrm{He}}
\newcommand{\aghe}{\textrm{Ag}\he}
\newcommand{\aghethree}{\textrm{Ag}\hethree}
\newcommand{\aghecol}{Ag--\hethree}
\newcommand{\aghehecol}{\aghethree--\hethree}
\newcommand{\vdw}{van der Waals molecule}
\newcommand{\vbar}{\ensuremath{\bar v}}

\newcommand{\longrightleftharpoons}{
	\mathop{\vcenter{\hbox{\ooalign{
		\raise 0.2ex\hbox{$\relbar\joinrel\rightharpoonup$}\crcr
		\lower 0.2ex\hbox{$\leftharpoondown\joinrel\relbar$}
	}}}}
}

\newcommand{\rates}[2]{\displaystyle
  \mathrel{\longrightleftharpoons^{#1\mathstrut}_{#2}}}

\newcommand{\sfrac}[2]{\ensuremath{{^{#1}\!/_{#2}}}}

\newcommand{\cmone}{\ensuremath{\:\mathrm{cm}^{-1}}}	
\newcommand{\cmthree}{\ensuremath{\mathrm{cm}^{-3}}}
\newcommand{\cmthrees}{\ensuremath{\mathrm{cm}^3/\mathrm{s}}}

\newcommand{\massaghe}{\ensuremath{2.92\:\mathrm{amu}}}	

\newcommand{\molconfig}{$X{^2\Sigma}^{+}$}	

\newcommand{\gtotal}{\gamma_\mathrm{tot}}
\newcommand{\gbound}{\gamma_\mathrm{bound}}
\newcommand{\gunbound}{\gamma_\mathrm{unbound}}

\newcommand{\threejm}[6]{ \left(\begin{array}{ccc} #1 & #3 & #5\\
                                              #2 & #4 & #6
                                \end{array}
                          \right)
}


\twocolumn[
  \begin{@twocolumnfalse}
\noindent\LARGE{\textbf{Formation and dynamics of van der Waals molecules in buffer-gas traps}}
\vspace{0.6cm}

\newcommand{\sucomma}{\textsuperscript{,}}
{
\renewcommand{\thefootnote}{\textit{\alph{footnote}}}
\noindent\large{\textbf{Nathan Brahms\footnotemark[1], Timur V. Tscherbul\footnotemark[2]\sucomma\footnotemark[3], Peng Zhang\footnotemark[3], Jacek K{\l}os\footnotemark[4], Robert C.\ Forrey\footnotemark[5], Yat Shan Au\footnotemark[2]\sucomma\footnotemark[6], H.R.\ Sadeghpour\footnotemark[3], A.\ Dalgarno\footnotemark[2]\sucomma\footnotemark[3], John M.\ Doyle\footnotemark[2]\sucomma\footnotemark[6], and Thad G.\ Walker\footnotemark[7]
}}\vspace{0.5cm}
}

\vspace{0.2in}

\noindent\normalsize{We show that weakly bound He-containing van der Waals molecules can be produced and magnetically trapped in buffer-gas cooling experiments, and provide a general model for the formation and dynamics of these molecules.  Our analysis shows that, at typical experimental parameters, thermodynamics favors the formation of van der Waals complexes composed of a helium atom bound to most open-shell atoms and molecules, and that complex formation occurs quickly enough to ensure chemical equilibrium.  For molecular pairs composed of a He atom and an $S$-state atom, the molecular spin is stable during formation, dissociation, and collisions, and thus these molecules can be magnetically trapped.  Collisional spin relaxations are too slow to affect trap lifetimes.  However, \hethree-containing complexes can change spin due to adiabatic crossings between trapped and untrapped Zeeman states, mediated by the anisotropic hyperfine interaction, causing trap loss.  We provide a detailed model for \aghethree\ molecules, using \textit{ab initio} calculation of Ag--He interaction potentials and spin interactions, quantum scattering theory, and direct Monte Carlo simulations to describe formation and spin relaxation in this system.  The calculated rate of spin-change agrees quantitatively with experimental observations, providing indirect evidence for molecular formation in buffer-gas-cooled magnetic traps.
}
\vspace{0.5cm}
 \end{@twocolumnfalse}
  ]

{
\renewcommand{\thefootnote}{\textit{\alph{footnote}}}
\footnotetext[1]{Department of Physics, University of California, Berkeley, California, United States; E-mail:\ \texttt{nbrahms@berkeley.edu}}
\footnotetext[2]{Harvard-MIT Center for Ultracold Atoms, Cambridge, Massachusetts, United States}
\footnotetext[3]{ITAMP, Harvard-Smithsonian Center for Astrophysics, Cambridge, Massachusetts, United States}
\footnotetext[4]{Department of Chemistry and Biochemistry, University of Maryland, College Park, Maryland, United States}
\footnotetext[5]{Department of Physics, Pennsylvania State University, Berks Campus, Reading, Pennsylvania, United States}
\footnotetext[6]{Department of Physics, Harvard University, Cambridge, Massachusetts, United States}
\footnotetext[7]{Department of Physics, University of Wisconsin, Madison, Wisconsin, United States}
}





The ability to cool and trap molecules holds great promise for new discoveries in chemistry and physics \cite{carr:NJP:review2009,softley:molphys:review,schnell:angewchem:review,ColdMoleculesBook}.  Cooled and trapped molecules yield long interaction times, allowing for precision measurements of molecular structure and interactions, tests of fundamental physics \cite{bohn:pra:edm,hudson:prl:edm}, and applications in quantum information science \cite{demille:prl:polarqi}.  Chemistry shows fundamentally different behavior for cold molecules \cite{krems:pccp:coldchem}, and can be highly controlled based on the kinetic energy, external fields \cite{sawyer:arxiv:collision}, and quantum states \cite{ospelkaus:science:controlledchem} of the reactants.

Experiments with cold molecules have thus far involved two classes of molecules: Feshbach molecules and deeply bound ground-state molecules.  Feshbach molecules are highly vibrationally excited molecules, bound near dissociation, which interact only weakly at long range, due to small dipole moments.  They are created by binding pairs of ultracold atoms using laser light (photoassociation), ac magnetic fields (rf association), or slowly varying dc magnetic fields (magnetoassociation).  By contrast, ground-state heteronuclear molecules often have substantial dipole moments and are immune to spontaneous decay and vibrational relaxation, making these molecules more promising for applications in quantum information processing \cite{demille:prl:polarqi} and quantum simulation \cite{andre:QS}.  These molecules are either cooled from high temperatures using techniques such as buffer-gas cooling or Stark deceleration, or are created via coherent deexcitation of Feshbach molecules.

This article introduces a third family to the hierarchy of trappable molecules -- van der Waals (vdW) complexes.  VdW molecules are bound solely by long-range dispersion interactions, leading to the weakest binding energies of any ground-state molecules, on the order of a wavenumber ($\sim$ one kelvin).  This is three orders of magnitude smaller than the binding energy of a typical ionically bound molecule.  The nuclei are bound at long range, and ground state electronic wavefunctions often differ from the constituent atomic wavefunctions only by small perturbations.  These unique characteristics have made vdW complexes an attractive platform for the study of chemical reaction dynamics \cite{balakrishnan,herschbach:Rg2,levy:NaAr,kalus:JCP99:QCT,gettys:JPC88:QCT} and surface interactions \cite{hutson:vdw:intermolecularForces:review}.  They also play a key role in nonlinear optical phenomena and decoherence of dense atomic and molecular gases \cite{happer:prl:MRg}. Once formed, the vdW molecules can decay via collision-induced dissociation \cite{parker:CID:jcp}, chemical exchange \cite{herschbach:Rg2,kalus:JCP99:QCT,gettys:JPC88:QCT} and electronic, vibrational, rotational, and Zeeman predissociation \cite{hutson:vdw:intermolecularForces:review,krems:vdw:magneticBreaking}. In vdW complexes with small binding energies (such as He-O), the Zeeman predissociation can be controlled with an external magnetic field \cite{krems:vdw:magneticBreaking}.

A particularly important class of vdW molecules, formed by an $S$-state metal atom (M) binding with a rare gas atom (Rg), has been the subject of extensive research for decades \cite{hutson:vdw:intermolecularForces:review,levy:science:vdw}.  These studies have used a variety of experimental techniques to produce and cool vdW molecules, including (1) supersonic expansions, (2) immersion in dense rare gas, (3) immersion in bulk liquid $^4$He, and (4) doping of superfluid $^4$He nanodroplets.  We present a brief overview of this earlier work below, with a particular emphasis on vdW complexes of alkali- and noble-metal atoms with Rg of relevance to the present work.

(1) In a typical molecular beam experiment, vdW clusters are produced in a supersonic expansion of M + Rg mixtures and probed via laser spectroscopy \cite{levy:science:vdw,mattison:KAr,bitar:NaNe,duncan:AgAr,jouvet:AgAr}. Examples include studies of NaNe and KAr via microwave spectroscopy \cite{mattison:KAr,bitar:NaNe},  CuAr \cite{duncan:CuKr}, AgAr \cite{duncan:AgAr}, and Ag$_2$Ar \cite{duncan:Ag2Ar} via two-photon ionization spectroscopy, AgAr via laser-induced fluorescence excitation spectroscopy \cite{jouvet:AgAr}, and AuRg via multiphoton ionization spectroscopy \cite{breckenridge:AuNe,breckenridge:AuAr}. These studies provide valuable spectroscopic information on fine and hyperfine interactions in vdW molecules in their ground \cite{mattison:KAr,bitar:NaNe} and excited \cite{breckenridge:AuNe,breckenridge:AuAr}  electronic states.

(2) VdW molecules also play an important role in the dynamics of species interacting with a dense rare gas vapor.  Spontaneous formation of vdW molecules plays a key role in the decoherence of alkali vapors in buffer-gas cells.  They were first observed in the spin relaxation of Rb due to the formation of RbAr and RbKr molecules \cite{bouchiat:alkalimolecules}.  Such molecules mediate efficient spin-exchange between alkali atoms and rare-gas nuclei \cite{walker:rmp} and have been shown to limit the precision of vapor-cell based atomic clocks \cite{happer:prl:MRg}.  Dense Rg vapors have also been used to spectroscopically observe MRg excimers \cite{takami:AgHe}.

(3)  For vdW molecules composed of a helium (He) atom dispersively bound to a moiety, binding energies are typically on the order of a wavenumber, corresponding to temperatures below one kelvin.  Due to their weak binding energies and vulnerability to collisions, He-containing vdW molecules are typically not observed in supersonic expansions, but can be produced by a number of alternative techniques such as immersing atoms in liquid $^4$He or dense $^4$He gas \cite{takahashi:AlkHe1, takami:AgHe}.  These studies have enabled the observation of vdW clusters formed from ions\cite{kojima:HHe+,koyanagi:BaRg2+} or excited state neutral atoms\cite{takami:condensation}.

(4) Yet another powerful experimental technique for studying vdW molecules is based on He nanodroplet spectroscopy.  Research has included spectroscopic investigations of the formation of RbHe exciplexes on the surface of a superfluid He droplet \cite{droppelmann:RbHe,mudrich:RbHe}, and of the effects of spin-orbit interactions on the formation of NaHe and KHe excimers upon optical excitation of alkali-doped nanodroplets \cite{reho:KHe,reho:NaHe}.  He nanodroplet spectroscopy can also give information on the character of M--Rg forces, showing, for example, that Ag atoms reside at the center of He nanodroplets\cite{toennies:AgHe,federmann:AgHe,mella:QMC}.

In this Article, we explore the formation and dynamics of He-containing van der Waals complexes in buffer-gas cooling experiments \cite{deCarvalho:bg:primer}.  This method is similar to the dense gas production method (2) above.  However, the production of van der Waals molecules in buffer-gas cooling experiments is favored by combining moderate Rg vapor densities with temperatures much colder than the vdW molecules' binding energy.  In contrast to the methods above, He-containing vdW molecules in such a system have relatively long chemical lifetimes (on the order of 10 to 100 milliseconds), and are favorably produced in their absolute rovibrational ground state.  These characteristics offer a unique laboratory for studying the dynamics of these molecules, including their formation processes, photochemistry, and scattering properties.  In addition, buffer-gas cooling experiments are readily integrated with external fields, especially magnetic traps.  A large number of He-containing vdW molecules are paramagnetic, raising the possibility for vdW molecule trapping with lifetimes up to hundreds of seconds, and perhaps even cooling into the ultracold regime.

We begin with a general model for the formation of He vdWs, describing the thermodynamics and chemical kinetics of the system.  We predict that a wide variety of reactants, including metal and nonmetal atoms, as well as molecular species, will readily bind in these systems.  We describe the collisional properties of these vdWs within magnetic traps.  We provide a detailed model for the case of \aghethree.  Using \textit{ab initio} models of the molecular structure and quantum collision calculations, we describe the formation and collision dynamics of \aghethree.  In a previous letter \cite{brahms:vdw}, we argued that the observed dynamics \cite{brahms:agTrapping} of Ag in a buffer-gas trap with \hethree\ provided indirect evidence for the formation of \aghethree\ molecules; here we use our theoretical model to present this argument in detail.  Finally, we discuss the possibility for spectroscopic observation of vdWs formed in buffer-gas cooling experiments, introducing a new technique enabled by the sensitivity of buffer-gas trapping to the collisional dynamics of vdW molecules.

\section{General model}
\label{ss:gen}

Buffer gas cooling experiments \cite{deCarvalho:bg:primer} operate by introducing a hot vapor of a target species $X$ (atom, molecule, or ion) into a  moderately dense, cryogenically cooled vapor, typically He.  After approximately 100 collisions, $X$ is cooled to the temperature of the coolant vapor, which can be as cold as 300 mK for \hethree, or 700 mK for \hefour.  The cold $X$ then diffuse through the He until they reach the walls of the cryogenic cell.

Without a magnetic trap, this method typically provides densities of $10^{12}~\cmthree$ for atomic species, or $10^{10}~\cmthree$ for molecular species.  The timescale for diffusive loss of the cooled species lies between a few and a few hundred milliseconds, proportional to the density of the buffer gas, which is controlled between a few $10^{14}$ and $10^{17}~\cmthree$.

Due to these low temperatures, the formation of van der Waals complexes can be thermodynamically favored.  We consider formation of vdWs due to inelastic three-body processes between a $X$ and two He atoms.  Such collisions can lead to pair formation via the process:
\begin{equation} \label{eqn:3BR}
X + \mathhe + \mathhe \rates{K}{D}  X\mathhe + \mathhe.
\end{equation}
Here $K$ and $D$ are the rate constants for three-body recombination and collision-induced dissociation.  We specifically ignore collisions involving multiple $X$ partners, as the density of He is typically 4 to 6 orders of magnitude larger than the $X$ density.

The pair formation and dissociation kinetics for this process are
\begin{equation}\label{eqn:molkin}
\dot{n}_{\xhe} = -\dot{n}_X = n_X/\tf -n_{\xhe}/\td,
\end{equation}
where $n_i$ denotes the density of species $i$, and the formation time $\tf$ and dissociation time $\td$ obey $1/\tf=K\nhe^2$ and $1/\td=D\,\nhe$.
If the timescale for pair formation and dissociation is fast compared to the lifetime of $X$ within the experiment, the density of pairs will come into thermal equilibrium with the free $X$ and He densities.
In thermal equilibrium, $\dot{n}_{\xhe} = 0$ implies $n_{\xhe} = \k(T) \nx \nhe$, where $\k(T) = K/D$ is the chemical equilibrium coefficient, derived from statistical mechanics \cite{reif}:
\begin{equation}
\k = \frac{\nxhe}{\nx \nhe} = \ldb^3\, \sum_i g_i e^{-\epsilon_i/ k_B\, T}.
\label{eqn:kappa:def}
\end{equation}
Here $k_B$ is Boltzmann's constant, $T$ is the temperature, and $\epsilon_i$ is the energy of molecular state $i$, having degeneracy $g_i$, where $\epsilon_i$ is zero at dissociation.  Note that bound states have negative $\epsilon_i$.  $\ldb$ is the thermal de Broglie wavelength of the complex, of reduced mass $\mu$, given by $\ldb = \sqrt{h^2/2\pi \mu k_B T}$. \eqn{eqn:kappa:def} shows that $\nxhe$ increases exponentially as the temperature is lowered, or as the binding energy of the molecule is increased.  A large number of vdW complexes have binding energies that are comparable to or greater than 1~K, and thus formation is thermodynamically favored in buffer gas cooling experiments. (Alkali-metal-He pairs \cite{mladenovic:prl:alkaliBinding} are a notable exception, having binding energies below 0.03~cm$^{-1}$.)  Table~\ref{table:dimerEnergies} gives a sampling of candidate $X$ species, showing binding energies and predicted population ratios $\frac{\nxhe}{\nx} = \k\nhe$ at standard temperatures and buffer gas densities.  

\eqn{eqn:molkin} implies that the timescale to reach thermal equilibrium is
\begin{equation}
\tau_\mathrm{eq}^{-1} = \tf^{-1} + \td^{-1} = D \nhe (1 + \kappa \nhe).
\end{equation}
An exact calculation of $D$ requires knowledge of the specific $X$He--He interaction potential.  However, an estimate can be obtained by assuming that dissociation occurs when the energy of the $X$He--He collision complex exceeds the molecular binding energy.  The fraction of collisions with energy greater than the binding energy is, in the low-temperature limit, $\sim e^{\epsilon_0/k_B T}$.  The formation rate is therefore estimated by letting $D = \sigma v_\mu e^{\epsilon_i/k_B T}$, where $\sigma$ is the elastic cross section and $v_\mu$ is the average relative collision velocity.  This gives
\begin{equation}
\tau_\mathrm{eq}^{-1} \sim \sigma \nhe v_\mu \left ( e^{\epsilon_i/k_B T} + g_i \ldb^3 \nhe \right ).
\end{equation}
Typical buffer gas cooling parameters are $\nhe=10^{16}~\cmthree$, with $T=300~\mathrm{mK}$ when using \hethree, or $T=700~\mathrm{mK}$ when using $\mathrm{^4He}$.  Using $\mu \approx 3~\textrm{or 4 amu}$ and a worst-case elastic cross section of $\sigma = 10^{-15}~\mathrm{cm^2}$ gives an equilibrium time $\tau_\mathrm{eq}\leq 4$~ms for pairs with $|\epsilon_0| < 1~\mathrm{cm^{-1}}$ and $\tau_\mathrm{eq} \leq 600$~ms for all values of $\epsilon_0$.  This timescale can be compared to the typical lifetime of a buffer gas trapped species, around 100 ms without a magnetic trap, and $\gtrsim 1$~s with a magnetic trap.  We therefore expect, in general, that the vdW pair density will reach thermal equilibrium in buffer gas cooling experiments.

\begin{table}[t]
\centering
\renewcommand{\thefootnote}{\textit{\alph{footnote}}}
\setcounter{footnote}{1}
\caption{Predicted ground-state energies $\epsilon_0$ and pair population ratios $\frac{\nxhe}{\nx} = \k\nhe$ of some species compatible with buffer gas cooling, for which $X$He internuclear potentials are available \cite{dimerTable1,cargnoni:mHePotentials,krems:prl:nons,maxwell:Bi:pra08}.  Molecules are assumed to be in their absolute rovibrational ground state, with interaction potentials taken from Refs.~\citenum{Cybulski2005:NHHe} (NH-He), \citenum{Groenenboom2003:CaHHe} (CaH-He), \citenum{krems:ybf:pra2007} (YbF-He), and \citenum{turpin:MnH} (MnH-He).
} 
\label{table:dimerEnergies}
\begin{revtabular}{ccr@{}lr@{}lr@{}lr@{}l}
		&       & \multicolumn{4}{c}{$X$$^3$He} & \multicolumn{4}{c}{$X$$^4$He} \\
$X$	& State & \mctwo{$-\epsilon_0$\footnotemark[1]} & \mctwo{$\frac{\nxhe}{\nx}$\footnotemark[2]} & \mctwo{$-\epsilon_0$\footnotemark[1]} & \mctwo{$\frac{\nxhe}{\nx}$\footnotemark[2]} \\
\midrule
Si	&	$^3P_{0}$	&	1&.49	&	0&.25		&	1&.95	&	0&.002	\\
Ge	&	$^3P_{0}$	&	1&.59	&	0&.41		&	2&.08	&	0&.003	\\
N	&	$^4S_{3/2}$	&	2&.13	&	6&.9		&	2&.85	&	0&.018	\\
P	&	$^4S_{3/2}$	&	2&.70	&	\mcc		&	3&.42	&	0&.046	\\
As	&	$^4S_{3/2}$	&	2&.76	&	\mcc		&	3&.49	&	0&.049	\\
Bi	&	$^4S_{3/2}$	&	28&.74	&	\mcc	&	33&.26&	\mcc		\\
O	&	$^3P_{2}$	&	4&.66	&	\mcc		&	5&.83	&	\mcc		\\
S	&	$^3P_{2}$	&	5&.05	&	\mcc		&	6&.34	&	\mcc		\\
Se	&	$^3P_{2}$	&	5&.21	&	\mcc		&	6&.50	&	\mcc		\\
F	&	$^2P_{3/2}$	&	2&.78	&	\mcc		&	3&.85	&	0&.13		\\
Cl	&	$^2P_{3/2}$	&	6&.02	&	\mcc		&	7&.48	&	\mcc		\\
Br	&	$^2P_{3/2}$	&	6&.31	&	\mcc		&	7&.75	&	\mcc		\\
I	&	$^2P_{3/2}$	&	7&.02	&	\mcc		&	8&.40	&	\mcc		\\
Li	&	$^2S_{1/2}$	&	\mctwo{$d$}	&	\mctwo{---}	&	0&.008&	7&$\times 10^{-5}$\\
Na	&  $^2S_{1/2}$	&  \mctwo{$d$}	&	\mctwo{---}	&	0&.03	&	5&$\times 10^{-5}$\\
Cu	&	$^2S_{1/2}$	&	0&.90	&	0&.015	&	1&.26 &	5&$\times 10^{-4}$\\
Ag	&  $^2S_{1/2}$	&	1&.40	&	0&.16		&	1&.85	&	0&.0016	\\
Au	&	$^2S_{1/2}$	&	4&.91	&	\mcc		&	5&.87	& 	6&.1		\\
\midrule
NH	&	$^3\Sigma^-$	&	3&.52	&	\mcc						&	4&.42	&	0&.41		\\
CaH &	$^2\Sigma^+$	&	0&.68	&	$5$&$\times 10^{-3}$	&	0&.96	&	$3$&$\times 10^{-4}$ \\
YbF &	$^2\Sigma^+$	&	4&.24	&	\mcc 						&	5&.57	&	\mcc \\
MnH &	$^7\Sigma^+$	&	0&.70	&	$6$&$\times 10^{-3}$	&	1&.01	&	$3$&$\times 10^{-4}$ 
\end{revtabular}
\\
\raggedright
\footnotemark[1]{
Energies in cm$^{-1}$, $1~\mathrm{cm}^{-1}\approx 1.4~\mathrm{K}$.
}
\\\footnotemark[2]{
Pair population ratios for $n_\text{He} = 3\times 10^{16}~\mathrm{cm}^{-3}$, at 300~mK for $X^3$He and at 700~mK for $X^4$He molecules, for the level with energy $\epsilon_0$.}
\\\footnotemark[3]{
At these parameters, pairs are subject to runaway clustering.  The equilibrium density can be chosen by raising $T$ or lowering $\nhe$.
}
\\\footnotemark[4]{
No bound states are predicted for Li$^3$He or Na$^3$He.
}
\end{table}

Until now, we have neglected the formation of larger vdW complexes.  These will form by processes similar to the pair formation process, via
\begin{equation}
X\mathhe_m + \mathhe + \mathhe \rates{K_m}{D_m}  X\mathhe_{m+1} + \mathhe.
\end{equation}
In thermal equilibrium, the density of clusters containing $m+1$ He atoms is related to the density of those having $m$ He atoms by the chemical equilibrium coefficient $\kappa_m$ for this process.  If we assume that all the $\kappa_m$ are approximately equal (reasonable for small $m$, where the He--He interactions are negligible), then the density of clusters with $m+1$ He atoms is $\approx n_X (\kappa n_\mathhe)^m$.  Higher-order clusters should therefore be favored once $\kappa n_\mathhe \gtrsim 1$.
See Table~\ref{table:dimerEnergies} for examples of atoms where clustering should occur.  For $-\epsilon_0$ on the order of a few cm$^{-1}$, clustering can be controlled by adjusting the cell temperature, with pairs favored for $k_B T \gtrsim -\epsilon_0 / 8$.  We posit that runaway clustering might be responsible for the heretofore unexplained rapid atom loss observed in experiments using 300 mK Au and Bi \cite{brahms:thesis,maxwell:thesis}.

\subsection{vdW complexes in magnetic traps}
\label{ss:gen:traps}

For paramagnetic species, both the density and lifetime of buffer-gas-cooled species can be significantly increased by the addition of a magnetic trapping field.  Such a trap is usually composed of a spherical quadrupole field, with a magnetic-field zero at the trap center, and a magnetic-field norm rising linearly to a few tesla at the trap edge.  A fraction (approximately half) of $X$ (the ``weak-field seekers'') will have magnetic-moment orientations such that their energy is minimized at the center of the trap.  Such a trap typically results in a two order of magnitude increase in density, and lifetimes up to tens of seconds.  However, in order to stay trapped, the magnetic orientation of the species with respect to the local field must be stable.  Spin-changing collisions with the buffer gas, in particular, will lead to relaxation of the magnetic orientation, and subsequently the relaxed particles will be lost from the trap.

We now show that, for the special case where $X$ is an $S$-state atom, the $X$He molecule will be spin-stable, and remain trapped, even through formation and collision processes.  Because the $X + \mathrm{He} + \mathrm{He}$ collision complex lacks any strong direct coupling between the spin orientation of $X$ and the degrees of freedom of the He atoms, we expect the spin orientation to be protected during both vdW association and dissociation.  A quantitative estimate for spin change during association and dissociation processes can be obtained by assuming this rate is similar to the spin-change rate in two-body $X + \mathrm{He}$ processes.  For species compatible with buffer-gas trapping \cite{deCarvalho:bg:primer}, this rate is typically negligible, on the order of $\lesssim 10^{-6}$ per formation or dissociation event.

Van der Waals pairs that have formed from a trapped $X$ and a He atom are, therefore, also trapped.  These pairs may, however, suffer spin-change at a rate faster than the unbound $X$.  Spin-change of vdW pairs has previously been studied in optical pumping experiments \cite{bouchiat:alkalimolecules}, in which a hot ($>300$~K) spin-polarized alkali vapor diffused in a Kr, Xe, or N$_2$ buffer gas.  In these experiments, vdW molecules were formed between the alkali metal and the buffer gas, and suffered spin-changing collisions with additional buffer gas.  In this process, the electron spin precesses internally due to either the spin-rotation or the hyperfine interactions; this precession can decohere during a collision.  

We consider these interactions using the molecular Hamiltonian
\begin{multline}\label{eqn:HMol}
\hat{H}_\text{mol} = \epsilon_{N}
  + A_X\bvec{I}_X\cdot \bvec{S} + \bvec{B}\cdot\left (2\mu_B\bvec{S} + \mu_X\bvec{I}_X + \mu_\mathrm{He}\bvec{I}_{\mathrm{He}} \right ) + \\ \gamma\bvec{N}\cdot \bvec{S} + A_\mathrm{He}\bvec{I}_\mathrm{He} \cdot \bvec{S}  + c \sqrt{\frac{8\pi}{15}} \sum_{q=-2} ^2 Y^*_{2,q}(\hat{r})[\bvec{I}_\mathrm{He}\otimes\bvec{S}]^{(2)}_q.
\end{multline}
Here $\epsilon_{N}$ is the rovibrational-electronic level energy and $\mathbf{B}$ is the magnetic field.  $\bvec{N}$ is the rotational angular momentum, $\bvec{S}$ is the electron spin, $\gamma$ is the spin-rotation constant, $\bvec{I}_X$ is the nuclear spin of $X$ with moment $\mu_X$, $A_X$ is the atomic hyperfine constant, and $\mu_B$ is the Bohr magneton.  The last two terms in \eqn{eqn:HMol} describe the isotropic and anisotropic hyperfine interaction of $\bvec{S}$ with a $^3$He nuclear spin $\bvec{I}_\mathrm{He}$.  We neglect both the small nuclear-spin-orbit interaction and the weak anisotropic part of the $\bvec{I}_X$--$\bvec{S}$ interaction.
The interaction parameters $\gamma$, $A_\mathrm{He}$, and $c$ can be estimated using the approximate methods contained in Ref.~\citenum{walker:rmp}.  

Now consider an interaction of the form $a \bvec{S}\cdot\bvec{J}$, where $\bvec{J}$ represents, e.g., $\bvec{N}$ or $\bvec{I}_\mathrm{He}$.
This interaction mixes the spin-polarized state with less-strongly trapped states.  Collisions can cause decoherence of this mixing, i.e. via angular momentum transfer, molecular dissociation, or nuclear exchange.  We overestimate the spin-change rate by making the gross approximation that all collisions cause decoherence with 100\% probability.  That is, we assume that collisions serve as projective measurements of $S_z$.

The probability of spin-change in a collision is now simply the overlap between the molecular eigenstate and more weakly trapped spin states.  For relatively large magnetic fields ($2 \mu_B B \gtrsim a$, which is the case for all but the central $\mu$m$^3$ of the trap), this overlap can be found using perturbation theory.  To first order, the interaction only causes an overlap with the $m_S = s\mathord{-}1$ state:
\begin{equation}
|\psi\rangle \sim \frac{|s,m_j\rangle + \frac{a \sqrt{2 s}}{4 \mu_B B} C_{j,m} |s\mathord-1,m_j\mathord+1\rangle }{1+\frac{a^2 s}{8 \mu_B^2 B^2}C_{j,m}^2}.
\end{equation}
Here $C_{j,m} = \sqrt{(j-m_j)(j+m_j+1)}$.  The probability $p_\mathrm{SC}$ that the spin relaxes is taken from an average over possible values of $m_j$, giving
\begin{equation}
p_\mathrm{SC} \sim \frac{a^2 s j(j+1)}{12 \mu_B^2 B^2}.
\label{eqn:psc}
\end{equation}
Finally, we average this probability over the magnetic-field distribution of an anti-Helmholtz quadrupole trap:
\begin{align}
\nonumber
\bar{p}_\mathrm{SC} &\sim \frac{1}{2} \left (\frac{2 \mu_B}{k_B T} \right )^3 \int B^2 e^{-2 \mu_B B/k_B T} p_\mathrm{SC}(B) dB, \\
	&\sim \frac{a^2 s (j^2 + j)}{6 k_B^2 T^2}.
\label{eqn:pscavg}
\end{align}
For typical internuclear separations of $\sim 10~a_0$, the hyperfine constants are $\sim h\times 1$~MHz.  For the contact hyperfine with $T\sim 0.3$~K, this becomes an average per-collision spin-change probability of a few $10^{-9}$.  With mean collision times on the order of a few microseconds, we find that this type of spin-changing collision will be too rare to significantly impact the trap lifetime.

A similar analysis as above can be applied to tensor interactions such as anisotropic hyperfine, with $|\Delta m|=2$ transitions also allowed at first order.  However, the results are of a similarly small magnitude.

We note that for ground-state molecules $N=0$.  Spin-rotation interactions, therefore, can only occur as a virtual coupling within collisions.  We explicitly consider this mechanism in our treatment of \aghehecol\ collisions below.  We find that the spin is similarly protected by the trap magnetic field, with a spin-change probability too small to play a role in trap loss.

Finally, the anisotropic hyperfine interaction can couple trapped ground states to untrapped excited rotational levels.  At trap magnetic fields where these levels cross, this coupling causes an avoided crossing, and molecules can adiabatically transfer between trapped and untrapped states.  In certain cases this can be the dominant loss process.  We detail this process in our treatment of \aghethree\ in \S\ref{ss:aghe:lz} below.

\section{A\ml{g}$\,^3\!$H\ml{e} molecules}

We now apply our analysis to the recently reported experimental work \cite{brahms:agTrapping} that studied silver (Ag) trapped using buffer-gas cooling with \hethree.  In this experiment, $\sim 10^{13}$ Ag atoms were cooled to temperatures between 300 and 700~mK using buffer-gas densities between $3\times 10^{15}$ and $10^{17}~\mathrm{cm}^{-3}$.  For all experimental parameters, exponential loss of the trapped atoms was observed.  By fitting the loss rate as a function of buffer-gas density at each temperature, the ratio of the rate of Ag spin-change to the elastic \aghecol\ collision rate was extracted.  This ratio displayed a strong empirical $T^{-6}$ temperature dependence.  In this section we apply quantum collision theory analysis to show that the observed spin-change rate could not result from standard \aghecol\ inelastic collisional process; we show that the formation and subsequent spin-change of \aghethree\ molecules quantitatively explains the observed spin-change rate.

\subsection{Molecular structure}\label{sec:MolecularStructure}
\label{ss:aghe:molstructure}

\begin{figure}[t] 
	\centering
		\includegraphics[width=0.95\linewidth]{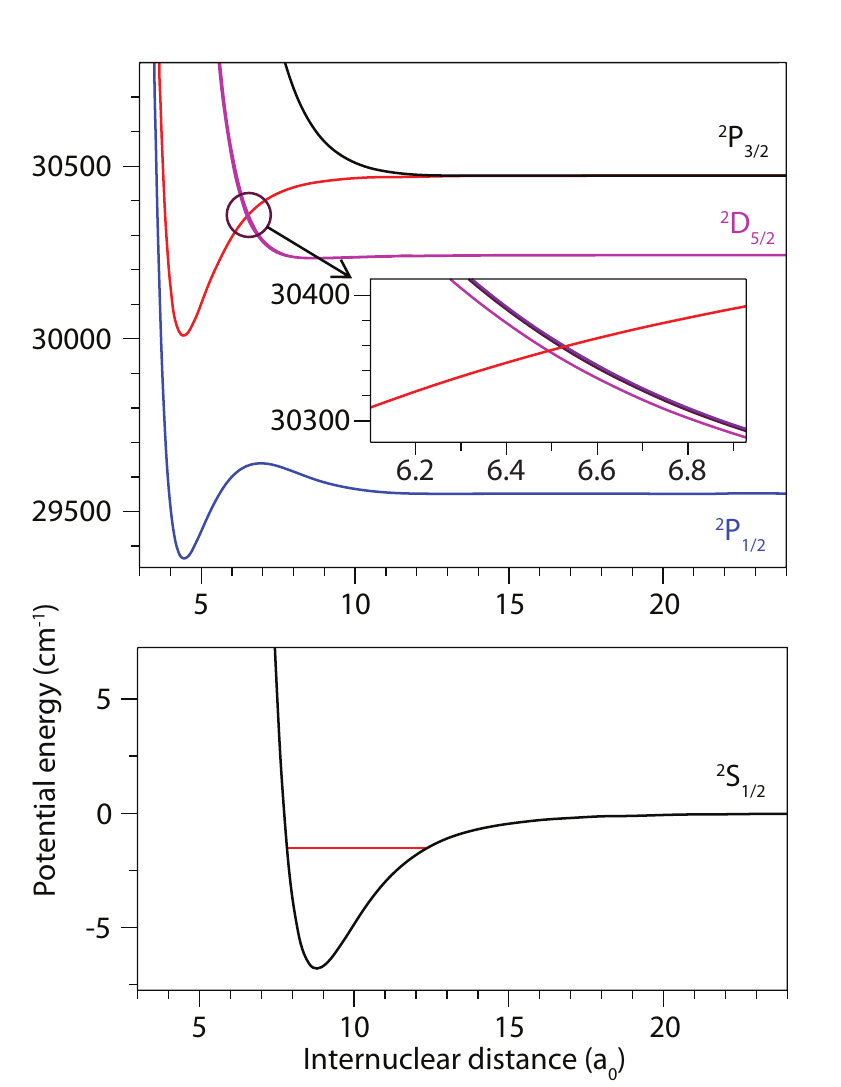}
	\caption{\textit{Ab initio} interaction potentials for the $X{^2}\Sigma$ (lower panel) and $\Omega=\sfrac 1 2$, $\sfrac 3 2$, and $\sfrac 5 2$ (upper panel) states of the AgHe complex as functions of the internuclear separation $r$.  The energy of the $N=0$ ground state is shown.  The excited-state potentials are calculated by diagonalizing the Hamiltonian matrix using the non-relativistic AgHe potentials of $\Sigma$, $\Pi$, and $\Delta$ symmetry computed in the present work (see text). The inset shows the region of (avoided) crossing  between the potential energy curves correlating with the $^2P_{3/2}$ and $^2D_{5/2}$ electronic states of Ag.}
	\label{fig:agHePotentials}
\end{figure}

We begin our analysis by constructing an \textit{ab initio} internuclear potential energy curve for Ag--He.  Potentials for this system have previously been constructed \cite{jakubek:agHe,cargnoni:mHePotentials} in the studies of the AgHe* exciplex.  An accurate potential has also been constructed \cite{cargnoni:2011} during the production of this article.   To construct our potential, shown in \fig{fig:agHePotentials}, we employed the partially spin-restricted coupled cluster method with single, double and perturbative triple excitations (RCCSD(T))~\cite{knowles:combo} as implemented in the \textsc{molpro} suite of programs~\cite{MOLPRO}.

The reference wave functions  for the electronic ground Ag($^2S$)-He and excited Ag($^2P$)-He and Ag($^2D$)-He complexes have been obtained from the restricted Hartree-Fock calculations (RHF). We employed the augmented, correlation-consistent basis set (aug-cc-pvqz) for the He atom~\cite{woon:94}. For the Ag atom, we used an effective core potential from the Stuttgart/Cologne group, which describes the first 28 electrons of the Ag atom (ECP28MCDHF)~\cite{figgen:05}, coupled with a pseudo-potential based on the aug-cc-pvqz-PP basis set of Peterson {\em et al.}~\cite{peterson:05} to describe the remaining 19 electrons.  This basis was additionally enhanced by using bond functions composed of $3s3p2d2f1g1h$ functions, with their origin half-way between the Ag and He atoms. The bond functions had the following exponents: $sp$, 0.9, 0.3, 0.1, $df$, 0.6, 0.2 and $gh$, 0.3.  The bond functions improve the description of the dispersion-governed interaction energies.  The interaction energy is corrected for the basis set superposition error by employing the counterpoise procedure of Boys and Bernardi~\cite{boys:70}. 

We monitored the T$_1$-diagnostic to be sure that the reference wave functions are mostly described by a single determinant. During coupled-cluster calculations the T$_1$ diagnostic was around 0.019 for Ag($^2S$)-He, 0.025 for Ag($^2P$)- and 0.022 for Ag($^2D$)-He, so we could apply a single-reference RHF/RCCSD(T) approach for all complexes.   

The positions and well depths of the potentials are characterized in Table~\ref{table:pes_features}.  The potentials are quite shallow, except that of the $A^2\Pi$ state originating from the $^2P$ term of the Ag atom. Our ground state X${{^2}\Sigma^+}$ Ag($^2S$)-He potential is similar to the one obtained with CCSD(T) method by Cargnoni {\em et al.}~\cite{cargnoni:mHePotentials}, while the older one from 1997 of Jakubek {\em et al.}~\cite{jakubek:agHe} has a deeper minimum and smaller equilibrium distance. Jakubek {\em et al.} used a high quality basis set, but the correlation treatment was done at the MP2 level, which tends to overestimate the binding energies.

Our results for the equilibrium distance and dissociation energy of the $^2\Sigma$ state are in good agreement with those by Tong {\it et al.} \cite{tong:09}, who used a similar basis set and the {\it ab initio} method. The value  $r_e=8.67$ obtained by Gardner {\it et al.} at the RCCSD(T)/d-awCV$\infty$Z level of theory is smaller and their dissociation energy is higher than both our result and that of Cargnoni \textit{et al.}.

The $A{^2\Sigma}$ state correlating to the $^2P$ term is practically repulsive; we find a very shallow minimum approximately 1~cm$^{-1}$ deep located very far from the Ag nucleus, at around 15 $a_0$. The $A{^2\Pi}$ potential of the AgHe* exciplex exhibits the deepest minimum relative to its asymptotic limit. We determined in our RCCSD(T) calculations that the well depth of this potential is approximately 464~cm$^{-1}$. Due to the effect of the bond functions our minimum is deeper than the one of Jakubek {\em et al.}\cite{jakubek:agHe} and the one of Cargnoni \textit{et al}. Cargnoni and Mella have published most recently new calculations \cite{cargnoni:2011} of the Ag($^2P$)-He potential using different levels of correlations in CISDT calculations, obtaining a well depth for the $A{^2\Pi}$ potential of 272 cm$^{-1}$.

\begin{table}
\renewcommand{\thefootnote}{\textit{\alph{footnote}}}
\centering
\caption{Equilibrium distances and well depths for non-relativistic Ag($^2S$)-He, Ag($^2P$)-He and Ag($^2D$)-He complexes. The minima are reported with respect to the asymptote of each electronic term of the Ag atom.}
\label{table:pes_features}
\begin{revtabular}{ccccc}
Complex	& State & $R_e~(a_0)$ & $D_e~(\mathrm{cm}^{-1})$ & Ref. \\
\midrule
Ag($^2S$)-He	&	$X{^2\Sigma^+}$	&	$8.80$	&	$6.80$	& Present	\\
              &               & $8.78$  & $6.81$  & \citenum{tong:09} \\ 
              &               & $8.67$  & $7.50$  & \citenum{gardner:agHe:10} \\              
							&               & $8.72$	& $7.22$  & \citenum{cargnoni:mHePotentials} \\   
              &               & $8.35$  & $11.3$  & \citenum{jakubek:agHe} \\ 
\midrule					
	Ag($^2P$)-He	&$A{^2\Sigma}$	&	$14.91$	&	$0.945$ & Present	\\
	            & $A{^2\Pi}$	    &	$4.42$	&	$463.6$ & \\  
	            &               & $5.16$  & $349.9$ & \citenum{jakubek:agHe} \\
	            &               & $4.76$  & $272.0$ & \citenum{cargnoni:2011} \\
\midrule
	Ag($^2D$)-He&$B{^2\Sigma}$	  &	$8.77$	&	$8.531$ & Present	\\
	            &$B{^2\Pi}$	      &	$8.74$	&	$8.673$ & \\
	            &$B{^2\Delta}$	  &	$8.67$	&	$8.942$ &
\end{revtabular}
\end{table}

The binding energies of vdW molecules containing $S$-state atoms were calculated by solving the one-dimensional Schr{\"o}dinger equation using the DVR method \cite{colbert:dvr:92}, yielding the rovibrational energy levels $\epsilon_{vN}$ and wavefunctions $\psi_{vN}(r)$. For vdW molecules formed by $P$-state atoms, both $V_\Sigma$ and $V_\Pi$ potentials were included in bound-state calculations and the variation of the spin-orbit splitting with $r$ was neglected \cite{carrington:jcp:94}. The binding energies of atom-molecule vdW complexes were calculated using the variational method of Ref. \citenum{grinev:jcp03} assuming the validity of the rigid-rotor approximation for rotational energy levels of the monomer. 

\begin{figure}[t] 
	\centering
		\includegraphics[width=0.95\linewidth]{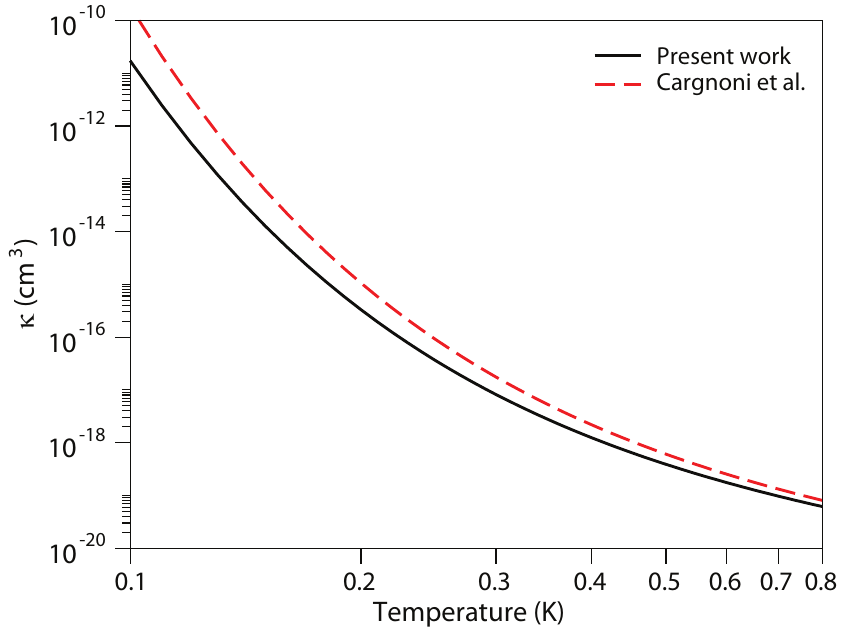}
	\caption{Chemical equilibrium constants for AgHe calculated as functions of temperature using the RCCSD(T) interaction potential computed in this work (full line) and the RCCSDT interaction potential (dashed line) from Ref.~\citenum{cargnoni:mHePotentials}.}
	\label{fig:chemEq}
\end{figure}

The AgHe molecular potential supports one vibrational bound state, with rotational quantum numbers $N = 0, 1, 2$ at energies $\epsilon_N = -1.40, -1.04, -0.37~\mathrm{cm}^{-1}$, shown in Fig.~\ref{fig:agHePotentials}.  We also note the existence of a quasibound $N=3$ state at $\epsilon_{3} = 0.48~\mathrm{cm}^{-1}$, with a calculated lifetime of $\approx 1~\mathrm{ns}$.  The predicted chemical equilibrium coefficient for \aghethree\ populations is shown in \fig{fig:chemEq}, alongside the prediction using the potentials of Ref.~\citenum{cargnoni:mHePotentials}. We observe that $\kappa(T)$ is very sensitive to fine details of the {\it ab initio} interaction potentials. At $T=0.5$~K, a 10$\%$ in the binding energy (from 1.40 to 1.54 cm$^{-1}$) leads to a 50$\%$ increase in $\kappa$.  Measurements of molecular populations may therefore serve as precise tests of intermolecular interactions.

\begin{figure}[t] 
	\centering
		\includegraphics[width=0.95\linewidth]{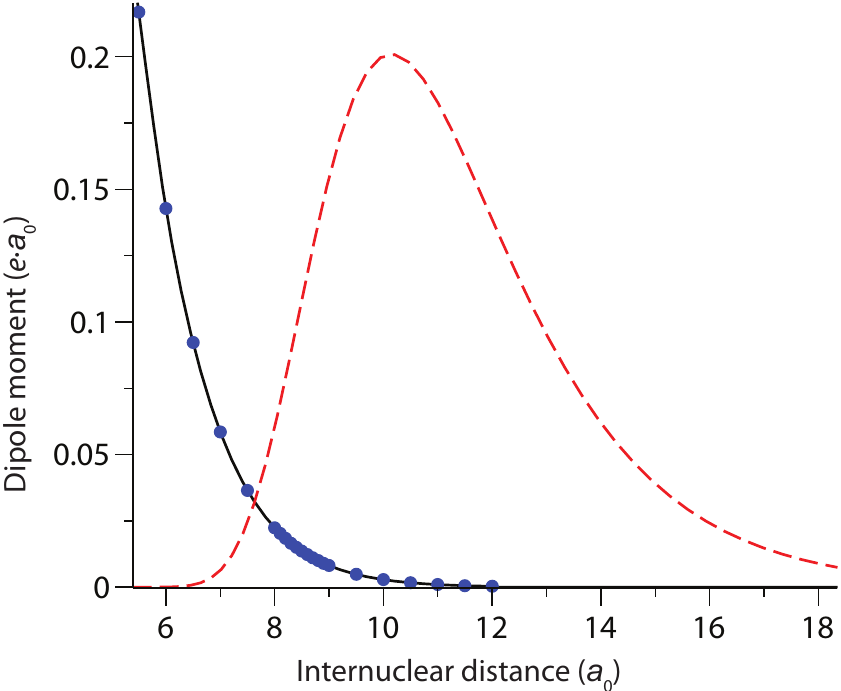}
	\caption{Ab initio dipole moment of the AgHe($^2\Sigma_{1/2}$) complex as a function of $r$ (circles). The ground-state rovibrational wavefunction of AgHe is superimposed on the plot (dashed line).}
	\label{fig:agHeDipole}
\end{figure}

\begin{figure}[t] 
	\centering
		\includegraphics[width=0.95\linewidth]{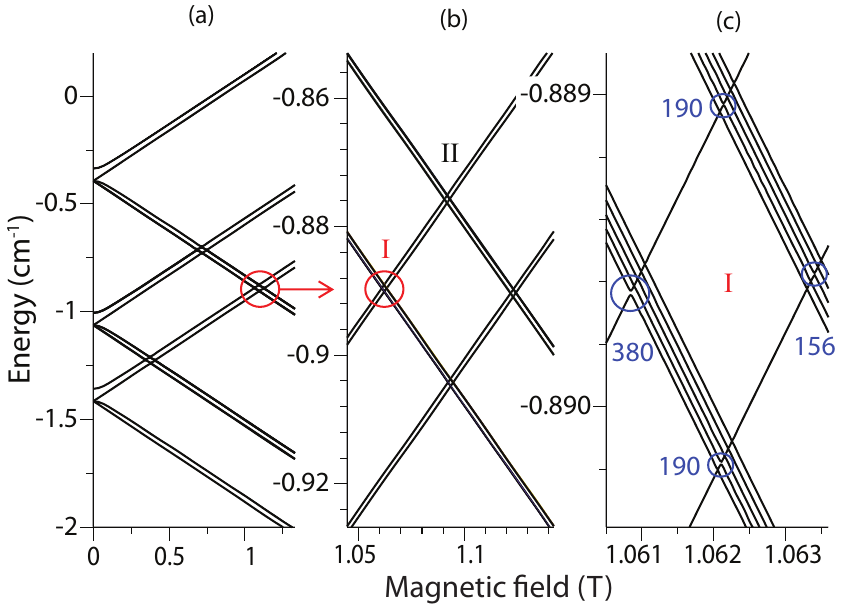}
	\caption{Zeeman energy levels of the AgHe molecule calculated by diagonalizing the Hamiltonian of \eqn{eqn:HMol}. The avoided crossing between the $N = 0$ and $N = 2$ rotational levels encircled in panel (a) is sequentially magnified in panels (b) and (c). The magnitudes of the splittings are given in kHz in panel (c).}
	\label{fig:agHeZeeman}
\end{figure}

The {\it ab initio} permanent electric dipole moment for the ground electronic state of AgHe is shown in \fig{fig:agHeDipole}. We observe that the molecule is slightly polar, having the expectation value of the dipole moment of $\langle \psi_{00}(r)|d(r)|\psi_{00}(r)\rangle = 0.004~e\cdot a_0$ in its ground state.  The Zeeman spectrum of \aghethree, consisting of identical hyperfine manifolds for each rotational level, is shown in \fig{fig:agHeZeeman}. 

To calculate the $r$-dependent isotropic and anisotropic hyperfine interaction constants $A_\text{He}(r)$ and $c(r)$, we employed quasirelativistic density functional theory (DFT) using the perturbatively corrected double hybrid functional B2PLYP \cite{hyperfine:Abinitio1}, which combines the virtues of DFT and second-order perturbation theory to improve the description of the electron correlation. Relativistic effects have been taken into account by the zeroth order regular approximation  (ZORA) \cite{hyperfine:Abinitio2,hyperfine:Abinitio3} of the Dirac equation, which has shown good performance for HFCs on heavy elements \cite{hyperfine:Abinitio4,hyperfine:Abinitio5}. A fully uncontracted and modified WTBS \cite{hyperfine:Abinitio6} basis ($31s21p19d7f4g$) was used for the Ag atom, where the $f$ and $g$ functions were taken from the $d$ functions, and a set of $spdf$ diffuse functions were added by the even-tempered manner, and two extra tight $s$ functions were augmented by multiplying the largest $s$ exponent of the parent basis by a factor of 3. For the He atom, a fully uncontracted ($12s7p4d3f2g$) basis \cite{hyperfine:Abinitio7} was adopted. The close agreement between the experimental hyperfine constant of the Ag atom, $A_\mathrm{Ag}/h=-1713$ MHz \cite{hyperfine:Abinitio9}, and the computed asymptotic value of $-1694$~MHz validates the current approach. 

For the spin-rotation parameter $\gamma(r)$, we use the perturbative result from Ref.~\citenum{walker:pra1997:spinRotation}:
\begin{equation}\label{eqn:gamma}
\gamma(r) = \frac{2 \hbar^6 a^2}{3 m_e^2 \mu r^2} \frac{\Delta_\text{SO}}{\Delta_{SP}^3}  \phi_S^2(r) \phi_P^2 (r),
\end{equation}
where $\Delta_\text{SO}/hc=920.642$ cm$^{-1}$ is the spin-orbit splitting of the lowest excited $^2P$ term of Ag, $\Delta_{SP}/hc=30165.8$ cm$^{-1}$ is the splitting between the $^2S$ and $^2P$ terms, $a=1.1784~a_0$ is the $s$-wave scattering length for electron-He collisions \cite{saha:scattering_length}, and $m_e$ is the electron mass.  The radial wavefunctions, normalized as $\int |\phi(r)|^2 r^2 dr = 1$, are taken from the  Hartree Fock of Ref.~\citenum{roetti:rhf} for the $5s$ state and calculated using the quantum defect method\cite{brahms:thesis} for the $5p$ state.

\begin{figure} 
	\centering
		\includegraphics[width=0.95\linewidth]{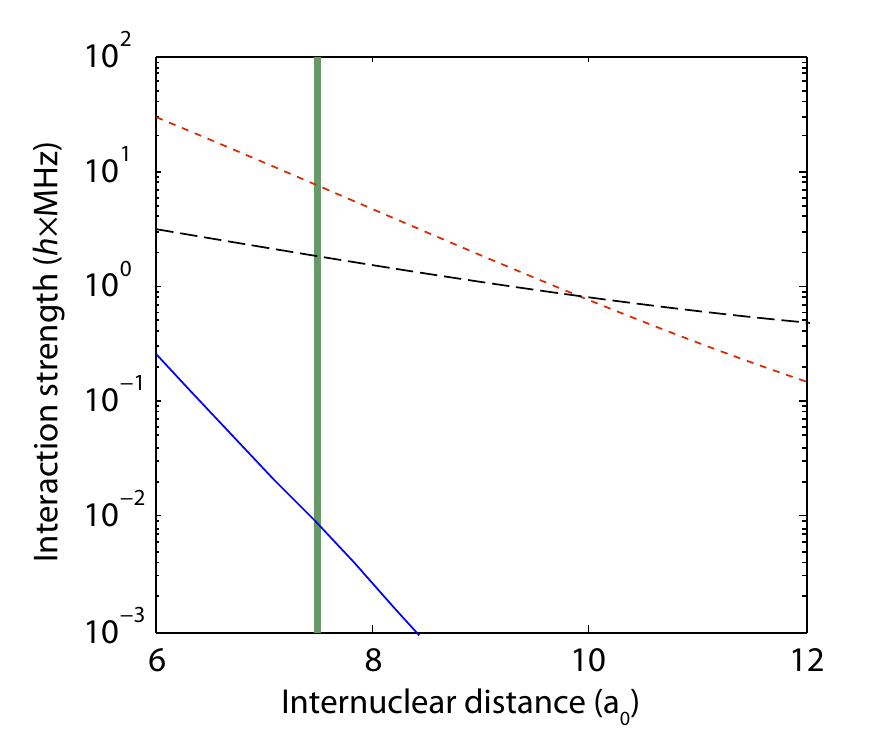}
	\caption{Spin-coupled interactions in the AgHe complex as a function of the internuclear separation: isotropic hyperfine interaction $A_\mathrm{He}$ (red dotted line), anisotropic hyperfine interaction $c$ (black dashed line), and unscaled ($f_s = 0$) spin-rotation interaction $\gamma$ (blue solid line).  The zero-energy classical turning point of the AgHe potential is shown by the green vertical line.  Note the rapid falloff of the spin-rotation interaction with $r$.}
	\label{fig:agHeInteractions}
\end{figure}

These interactions, as functions of nuclear distance, are shown in \fig{fig:agHeInteractions}.  Their values, averaged over the $N=0$ nuclear wavefunction, are $A_\mathrm{He} = -h\times 0.9$~MHz, $c = -h\times 1.04$~MHz, and $\gamma = h\times 180$~Hz.

\subsection{\aghecol\ collisions}
\label{ss:aghe:atomcol}

\begin{figure} 
	\centering
		\includegraphics[width=0.95\linewidth]{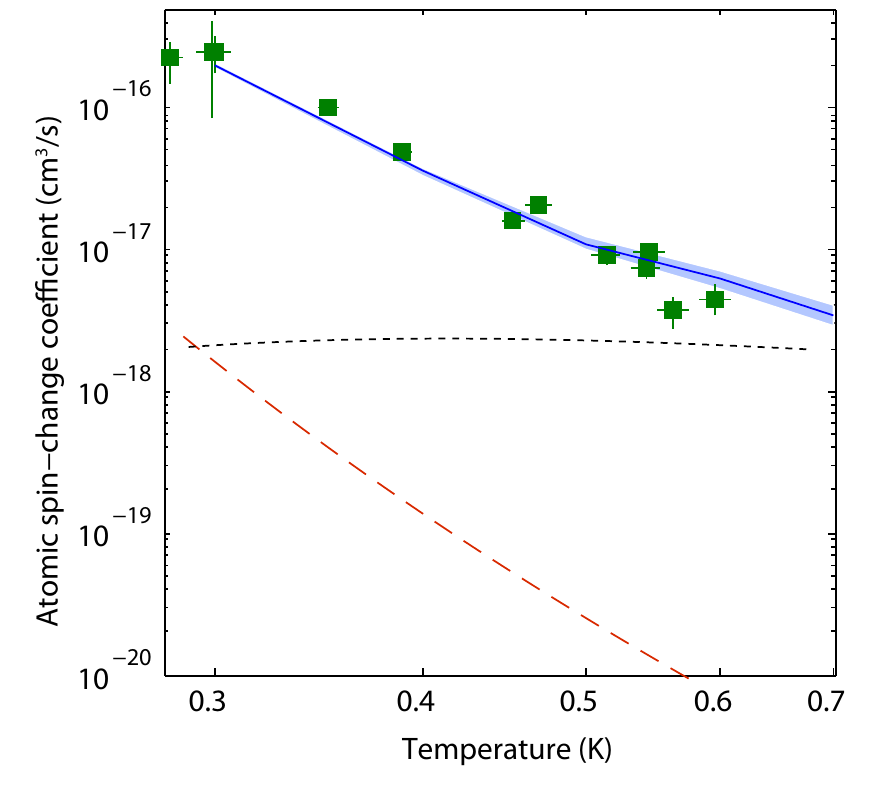}
	\caption{Spin-change rate of buffer-gas trapped Ag for $\nhe = 3\times 10^{16}~\mathrm{cm}^{-3}$.  The experimental data (green squares) disagree with the calculated maximum contributions from \aghecol\ (black dotted line) and \aghehecol\ collisional spin relaxation (red dashed line).  The data are well explained by Monte Carlo simulations of loss due to adiabatic following at the anisotropic-hyperfine-induced avoided crossings (blue line, shaded area indicates 68\% confidence interval).}
	\label{fig:agHeHeSpinRate}
\end{figure}

\begin{figure}[t] 
	\centering
		\includegraphics[width=0.95\linewidth]{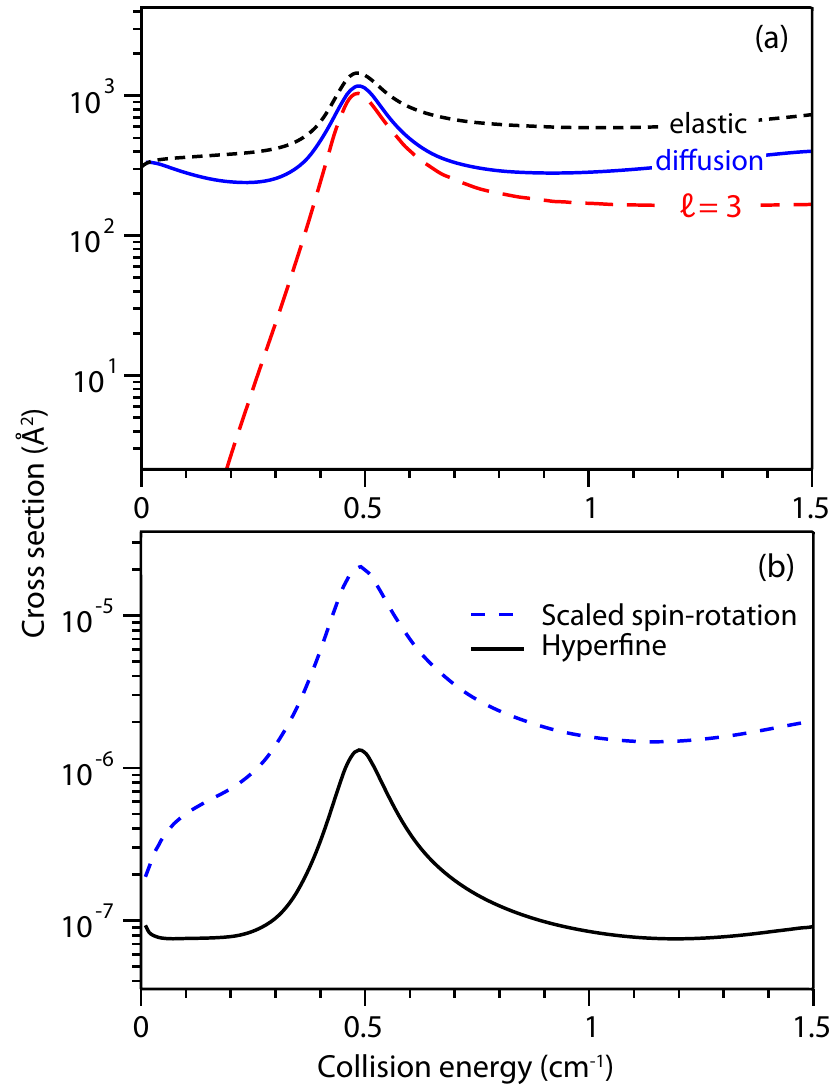}
	\caption{Elastic (upper panel) and spin relaxation (lower panel) cross sections for \aghecol\ collisions as functions of collision energy.  The solid line in the lower panel shows the contribution due to hyperfine interactions, while the dashed line shows the contribution due to an inflated (by a factor of 9,000) spin-rotation interaction.}
	\label{fig:agHeCol}
\end{figure}

We now calculate the spin-change rate $g_\mathrm{SC}$ of buffer-gas trapped Ag due to two-body \aghecol\ collisions.  We first calculate the \aghecol\ elastic and diffusion cross sections $\sigma_d$ using the {\it ab initio} AgHe potential calculated in this work (Sec. \ref{ss:aghe:molstructure}) by numerically integrating the Schr\"{o}dinger equation for collisional angular momentum up to $\ell = 5$ to produce the cross sections as a function of collision energy, shown in \fig{fig:agHeCol}(a).  The experimentally measured rate of atomic spin-change can be extracted from the measured elastic-inelastic ratio $\xi$ using $g_\mathrm{SC} = \xi \sigma_d v_\mu$.  The experimental spin-change rates are shown in \fig{fig:agHeHeSpinRate}.

To show that two-body atomic collisions cannot account for the measured thermal behavior of the spin-change rate, we performed quantum collision calculations of spin exchange and spin relaxation rates in Ag($^2S$)--$^3$He collisions using the same quantum scattering approach as developed earlier for the alkali-metal atoms \cite{tscherbul:AlkHe:2008,tscherbul:AlkHe:2009}. The Hamiltonian of the Ag($^{2}S$)-$^3$He collision complex in an external magnetic field $\mathbf{B}$ is similar in form to that given by \eqn{eqn:HMol}
\begin{multline}\label{eqn:Hcoll}
\hat{H}_\text{col} = -\frac{\hbar^2}{2\mu r^2}\frac{\partial^2}{\partial r^2}r + \frac{\mathbf{\ell}^2}{2\mu r^2} 
  + A_\mathrm{Ag}\bvec{I}_\mathrm{Ag}\cdot \bvec{S} \\
  + \bvec{B}\cdot\left (2\mu_B\bvec{S} + \mu_\mathrm{Ag}\bvec{I}_\mathrm{Ag} + \mu_\mathrm{He}\bvec{I}_{\mathrm{He}} \right ) + \gamma(r)\mathbf{\ell}\cdot \bvec{S} \\
  + A_\mathrm{He}(r)\bvec{I}_\mathrm{He} \cdot \bvec{S}  + c(r) \sqrt{\frac{8\pi}{15}} \sum_{q=-2} ^2 Y^*_{2,q}(\hat{r})[\bvec{I}_\mathrm{He}\otimes\bvec{S}]^{(2)}_q.
\end{multline}
where $r$ is the interatomic separation, $\mu$ is the reduced mass of $^{107}$Ag--$^3$He and $\mathbf{\ell}$ is the orbital angular momentum for the collision (replacing the rotational angular momentum $\bvec{N}$ in \eqn{eqn:HMol}). Unlike the molecular Hamiltonian \eqn{eqn:HMol}, the Hamiltonian given by \eqn{eqn:Hcoll} depends explicitly on $r$. The three last terms in \eqn{eqn:Hcoll} describe, respectively, the spin-rotation, isotropic hyperfine, and anisotropic hyperfine interactions.  We explicitly ignore the small effect of the anisotropic modification of the Ag hyperfine interaction due to the interaction with the He atom.
  
Having parametrized the Hamiltonian of \eqn{eqn:Hcoll}, we solve the scattering problem by expanding the wavefunction of the \aghe\ complex in the fully uncoupled basis:
\begin{equation}\label{UncoupledBasis}
|S m_S\rangle |I_{X} m_{I_X} \rangle |I_\mathrm{He} m_{I\mathrm{He}}\rangle |l m_l \rangle.
\end{equation}
where $m_S$, $m_{I_X}$, and $m_{I\mathrm{He}}$ are the projections of $S$, $I_\mathrm{Ag}$, and $I_\text{He}$ on the magnetic field axis. The system of close-coupled Schr{\"o}dinger equations   for the radial wavefunctions is solved for fixed values of the collision energy and magnetic field. 
The scattering matrix is evaluated in the basis \cite{tscherbul:AlkHe:2009} which diagonalizes the asymptotic Hamiltonian given by the second and the third terms on the right-hand side of \eqn{eqn:Hcoll}, which yields the probabilities for collision-induced transitions between different hyperfine states of Ag. We consider collisions of Ag atoms initially in the low-field-seeking hyperfine state $|F=0,m_F=0\rangle$.

\fig{fig:agHeCol}(a) shows the elastic cross section for \aghecol\  collisions plotted as a function of the collision energy. The cross section displays a pronounced peak near $0.5$ cm$^{-1}$, which is due to an $\ell=3$ shape resonance in the incident collision channel. 
The calculated spin-relaxation cross section is shown in \fig{fig:agHeCol}(b), and is, not surprisingly, dominated by contributions from the hyperfine interactions. 

By averaging the cross sections shown in \fig{fig:agHeCol}(b) with a Maxwell-Boltzmann distribution of collision energies, we obtain the inelastic \aghecol\ collision rates as functions of temperature, shown in \fig{fig:agHeHeSpinRate}.  The spin-relaxation rate, calculated for the average trap field of $B = 0.4$ T\footnote{The \aghecol\ rate varies only slightly with field.}, increases only slowly with temperature.  The rate remains small in absolute magnitude compared to the molecular spin relaxation rates considered in the next section.  We also perform the calculation with a grossly exaggerated spin-rotation parameter (by a factor of 9,000), inflated to match the experimental measurement at 700~mK.  The disagreement of this calculation with measurement at low temperatures indicates that, even if the magnitude of the perturbative result is incorrect, \aghecol\ collisions are not responsible for the observed spin-change rate.  Furthermore, the observed loss was exponential in time, rather than following the $1/(a+t)$ profile that would result if Ag--Ag collisions were responsible for the Ag spin relaxation.  We therefore conclude that single atomic collisions can not be responsible for the experimentally observed behavior.

In contrast, we see a marked similarity between the thermal behavior of the measured relaxation rate and the temperature dependence of the \aghe\ chemical equilibrium coefficient (see \fig{fig:chemEq}).  This agreement strongly suggests that molecules form within the trap, and that it is these molecules which suffer spin relaxation, thereby causing the observed Ag trap loss.  We begin our treatment of the molecular dynamics by calculating the molecular formation kinetics of \aghethree.

\subsection{Molecular formation}
\label{ss:aghe:formation}

\begin{figure}[t] 
	\centering
		\includegraphics[width=0.95\linewidth]{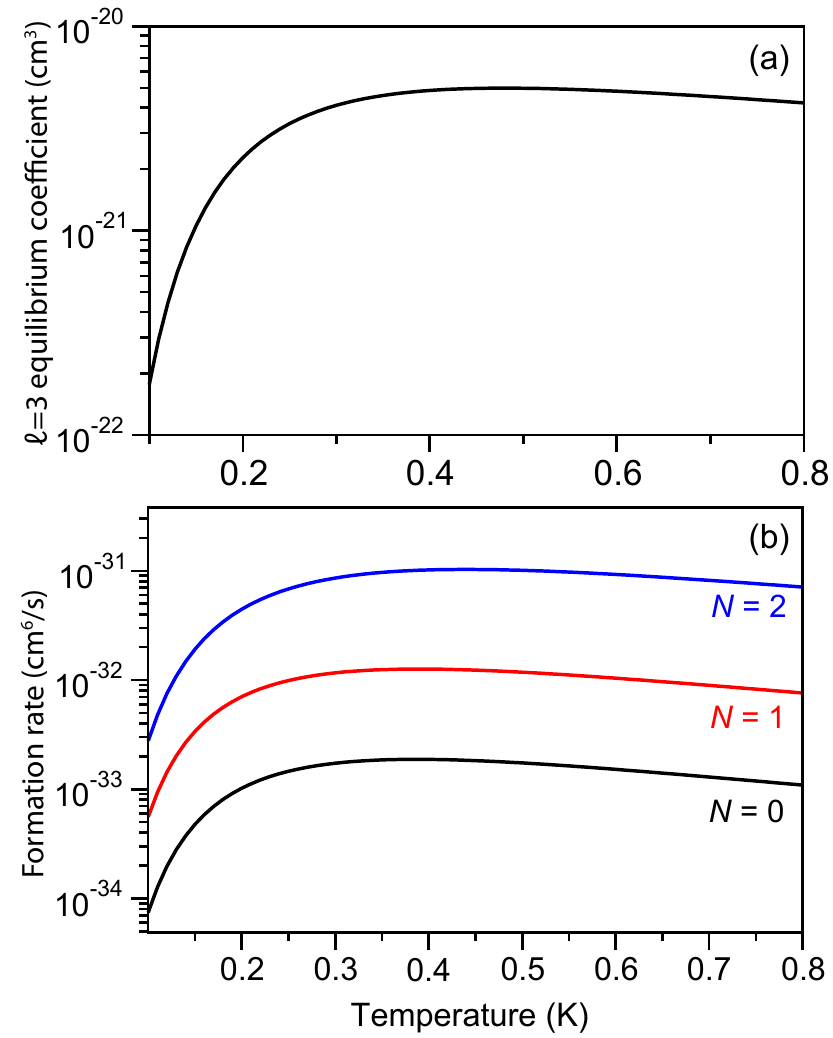}
	\caption{(a) The equilibrium constant for the formation of metastable AgHe*$(N=3)$ complex as a function of temperature. (b) Rate constants for three-body recombination Ag + He + He $\to$ AgHe($N'$) + He calculated using the RBC model as functions of temperature. Each curve is labeled by the final rotational state $N'$ of AgHe. }
	\label{fig:agHeFormation}
\end{figure}

We consider two mechanisms for the formation of \aghethree\ molecules.  The first is formation via the $\ell=3$ shape resonance shown in Fig.~\ref{fig:agHeCol}.  In this mechanism, ground state molecules form in a sequence of two-body collisions.  In the first collision, an Ag and a He atom form a quasibound pair.  During the lifetime of this pair, another He atom collides with the complex, causing rotational relaxation of the pair into a bound rotational level.  Additional collisions cause rotational relaxation into the rotational ground state:
\begin{align}
\mathrm{Ag} + \hethree &\longrightleftharpoons \aghethree^*(N\mathord{=}3), \\
\aghethree(N) + \hethree &\rates{g_{N, N'}}{g_{N', N}} \aghethree(N') + \hethree.
\end{align}
To calculate the formation rate of ground-state \aghethree\ pairs, we apply the resonant three-body recombination model developed by Roberts, Bernstein, and Curtiss (RBC) \cite{rbc:jcp:recombination} (also known as the Lindemann recombination mechanism).  Under this model, the rate coefficient for bound pair formation is simply the product of the equilibrium coefficient for quasibound pairs times the rotational relaxation rate coefficient:
\begin{equation}
K_r = \kappa(\epsilon_3) \sum_N g_{3,N} = 7 \ldb^3 e^{-\epsilon_3/ k_B\, T} \sum_N g_{3,N},
\end{equation}
where the factor of 7 arises from the degeneracy of the $N=3$ state.
The rotational relaxation rate coefficients are calculated using the atom-molecule collision theory described in the next section. For the state-to-state rotational relaxation rates from the $N=3$ quasibound state we find $g_{3,2} = 2.0\times 10^{-11},~ g_{3,1} = 2.4\times 10^{-12}$, and $g_{3,0} =3.5\times 10^{-13}~\mathrm{cm^3/s}$ at $T = 0.5$ K.  The calculated formation rates are shown vs.\ temperature in \fig{fig:agHeFormation}, and are dominated by resonant combination into the $N=2$ level.  The formation rate is between 0.8 and $1.0\times 10^{-31}~\mathrm{cm^6/s}$ for all temperatures in the experiment.  After formation in the $N=2$ level, rotational thermalization proceeds via additional rotational relaxation. The rate constants for rotational relaxation from the $N=2$ and $N=1$ rotational levels are similar to those from the $N=3$ level, so the timescale for rotational thermalization is fast (by a factor $\approx \kappa(\epsilon_3) \nhe$) compared to the molecular recombination rate, and $K_r$ therefore sets the timescale for resonant ground-state molecule formation.  Because $K_r \nhe^2 \leq 100~\mathrm{ms}$ is much less than the trap lifetime $\tau_\mathrm{trap} \geq 400~\mathrm{ms}$ for all values of $\nhe$ and $T$ used in the experiment, the molecular density, and hence the molecular spin-change dynamics, can be calculated assuming thermal equilibrium.

\begin{figure} 
	\centering
		\includegraphics[width=0.95\linewidth]{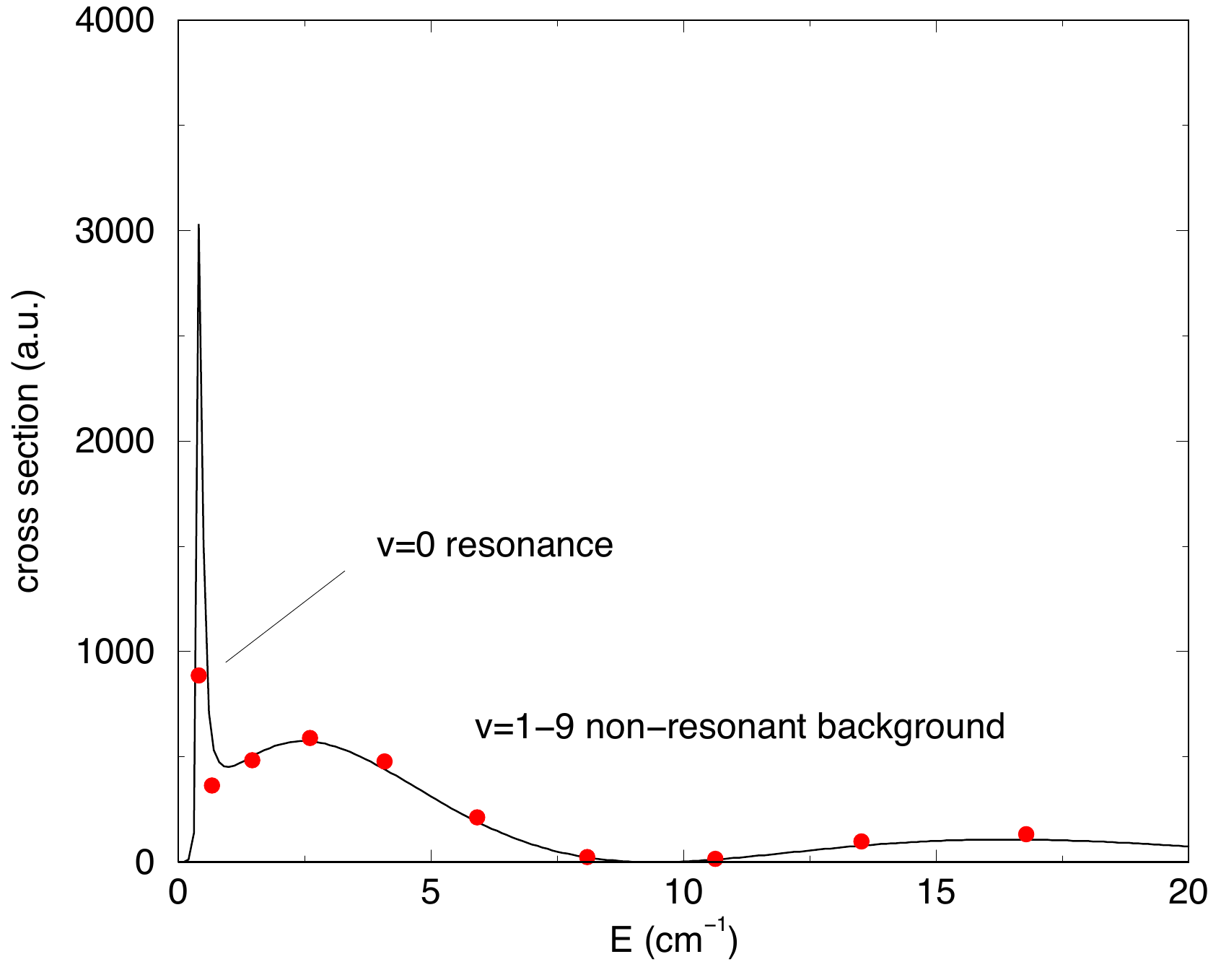}
	\caption{Elastic scattering cross section for Ag + $^3$He with $\ell=3$. The solid black curve was computed using numerical integration, and the red circles were computed using the Sturmian basis set representation. The $v=0$ eigenstate is associated with the resonance, and the $v>0$ eigenstates with the non-resonant background.}
	\label{fig:basis}
\end{figure}

The second formation mechanism is ``direct" formation via the three-body process
\begin{equation}
\mathrm{Ag} + \hethree + \hethree \rates{K_d}{D_d} \aghethree + \hethree. 
\end{equation}
An exact calculation of the formation rate via this mechanism lies outside the scope of this article.  However, the rate may be approximated by extending the sequential RBC orbiting resonance theory to include contributions from the non-resonant two-body continuum. Because the time delays of the non-resonant states are negligible, this approach gives pure third-order kinetics for all densities. We use the index $i$ to denote an unbound or quasibound initial state of AgHe and compute the recombination rate coefficient using
\begin{align}
K_r&=\ldb^3
\sum_{if}k_{if}\,g_i\,e^{-E_i/k_BT},
\end{align}
where
\begin{equation}
k_{if}=\sqrt{\frac{8}{\pi\mu k_B^3 T^3}} \int_0^\infty \sigma_{if}(E_T)e^{-E_T/k_BT}\,E_T\,dE_T.
\end{equation}
The energy $E_T=E-E_i$ is the translational energy in the $i$th channel, and $\sigma_{if}(E_T)$ is the collision cross section for transition to a bound final state. The energy $E_i$ is a positive energy eigenvalue of the diatomic Schr\"odinger equation in a Sturmian basis set representation, which may correspond to a resonance or to a discretized non-resonant contribution in a numerical quadrature of the continuum \cite{paolini:2011}.  We use a Sturmian basis set representation consisting of 100 Laguerre polynomial $L_n^{(2l+2)}$ functions of the form
\begin{equation}
\phi_{l,n}(r) = \sqrt{\frac{a_s n!}{(n+2l+2)!}}
\,(a_sr)^{l+1}e^{-a_sr/2}L_n^{(2l+2)}(a_sr),
\end{equation}
with a scale factor $a_s=10$. To assess the quality of this representation, the positive energy eigenstates were used to compute the Ag+He elastic scattering cross section for $\ell=3$. The results are shown in \fig{fig:basis} along with the exact results obtained from numerical integration. The lowest energy $v=0$ eigenstate is clearly associated with the resonance, whereas the $v>0$ eigenstates may be associated with the non-resonant background.

\begin{figure}[t] 
	\centering
		\includegraphics[width=0.95\linewidth]{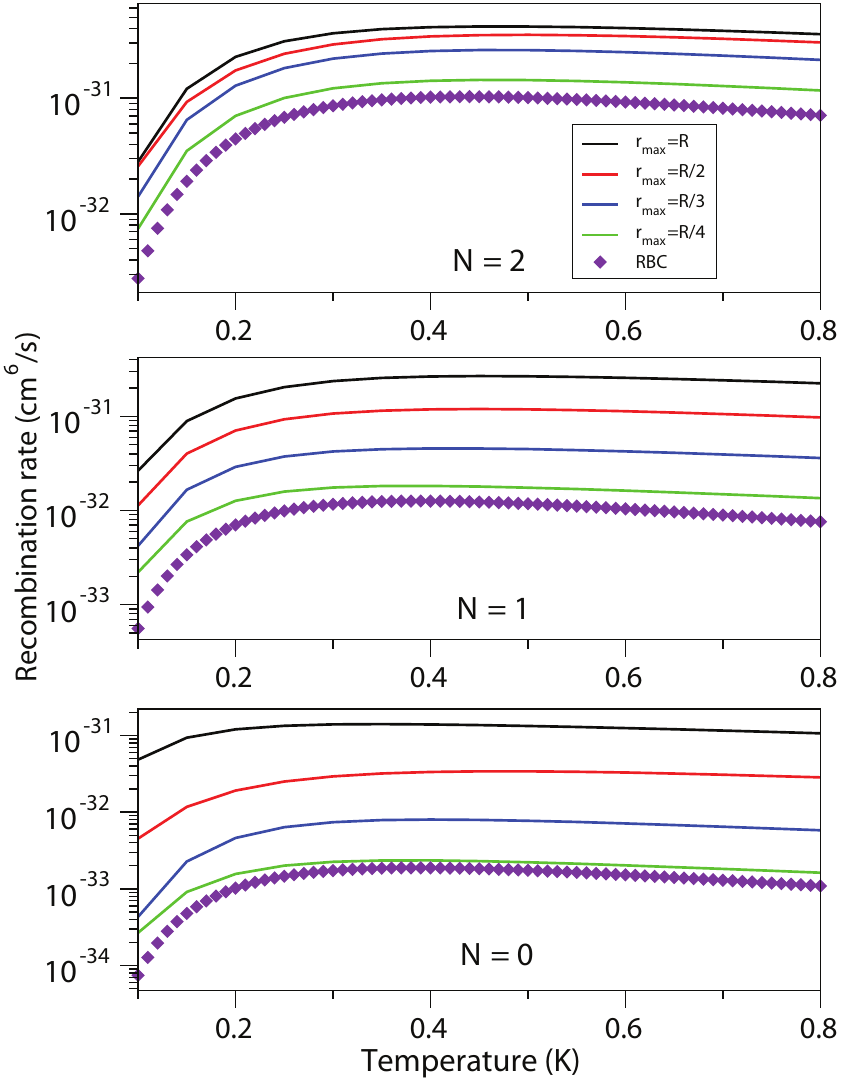}
	\caption{Three-body recombination rates for the formation of AgHe molecules in different rotational states ($N$) as functions of temperature: $N=2$ (upper panel), $N=1$ (middle panel), and $N=0$ (lower panel). Solid lines -- results calculated using the continuum discretization method (see text), symbols -- results obtained using the RBC model.}
	\label{fig:agHeRBC}
\end{figure}

The collision cross sections $\sigma_{if}$ are computed by solving a set of coupled channel (CC) equations using the general inelastic scattering program \textsc{molscat} \cite{molscat}.  The recombining He atom is assumed to be distinguishable from the colliding He atom and to be nearest to the Ag atom.  As discussed below, the interaction PES becomes more anisotropic with increasing $r$. We investigated the sensitivity of the CC calculations to this anistropy by setting $V(R, r, \theta)=V(R, r_{max}, \theta)$ for $r\geq r_{max}$ and
considering $r_{max}=R/n$ with $n=1-4$. The results are shown in Figure \ref{fig:agHeRBC}. Also shown are results of the $\ell=3$ resonant contribution using the rigid rotor approximation with $r=3~a_0$.
The CC calculations that include vibrational coupling were found to converge using $v_{max}=6$ and $j_{max}=6$ for a total of 49 basis functions. Convergence of the Legendre expansion 
\begin{equation}\label{eqn:LegendreExpansion}
V(R,r,\theta) = \sum_{\lambda=0}^{\lambda_\mathrm{max}} V_\lambda(R,r) P_\lambda (\cos\theta).
\end{equation}
was found for $n=3$ using $\lambda_{max}=6$. For consistency, these parameters were used for each of the CC calculations along with a matching distance
$R_{max}=100~a_0$, a maximum total angular momentum quantum number $J_{max}=10$, and a 20 point numerical integration over $\theta$.  The figure shows that the CC results with $n=4$ are similar to the RBC results using the rigid rotor approximation. The direct three-body recombination mechanism is essentially negligible in this case. The vibrational coupling to the non-resonant background is more substantial as $n$ decreases causing the recombination rate for each $N$ to increase. It is difficult to pinpoint precisely how much increase may be expected for the exact potential, however, the $n=1$ results provide a reasonable estimate.  The convergence of both the Legendre expansion and the basis set representation begins to break down as 
$n$ is reduced further, which suggests that a chemical exchange mechanism may be significant for this system. This possibility will be considered in a future study.

\subsection{\aghehecol\ collisions} 
\label{ss:aghe:molcol}

Once formed, the \aghethree\ molecules can undergo spin relaxation in collisions with $^3$He atoms, which convert low-field-seeking states to high-field-seeking states, leading to trap loss. In this section, we estimate the rate for spin-flipping \aghehecol\ collisions, and show that it is too small to account for the experimentally measured\cite{brahms:agTrapping} trap loss rates.

Low-temperature collisions involving vdW molecules may lead not only to inelastic spin relaxation, but also chemical exchange and three-body breakup. A proper theoretical description of these processes requires the use of hyperspherical coordinates \cite{parker:CID:jcp}, and is beyond the scope of this work. In order to estimate spin relaxation probabilities in \aghehecol\ collisions, we instead assume that:

\begin{enumerate}
\item The energy spectrum of \aghethree\ is described by the rigid-rotor Hamiltonian of \eqn{eqn:HMol}. This approximation is justified for the first three rotational levels ($N=0-2$). The $N = 3$ level corresponds to the long-lived shape resonance shown in \fig{fig:agHeCol}, and can thus be treated in the same manner as the (truly bound) lower rotational states. 

\item The contributions to spin relaxation collisions due to chemical exchange ($\mathrm{AgHe} + \mathrm{He}' \to \mathrm{AgHe}' + \mathrm{He}$) and collision-induced dissociation ($\text{AgHe + He} \to \text{Ag + He + He}$) can be neglected. Because of the short-range nature of these processes, they might result in a more efficient spin relaxation than inelastic collisions alone. Therefore, our estimates for spin relaxation are best thought of as lower bounds to true molecular spin relaxation rates.

\item The contributions of the hyperfine interactions to the loss rate can be described by the perturbative result in \eqn{eqn:psc} (in fact, this equation should overestimate their contribution).  We will therefore only perform a quantum collision calculation of the contribution from the spin-rotation interaction.  In order to place an upper limit on this contribution, we will use the scaled value of \S\ref{ss:aghe:atomcol}.

\item The interaction potential for AgHe-He is the sum of pairwise interaction potentials for Ag-He and He-He evaluated at a fixed AgHe distance of 3.0 $a_0$. We choose this value in order to ensure the convergence of the Legendre expansion of the AgHe$_2$ interaction potential (see below).
\end{enumerate}

Under these assumptions, the Hamiltonian of the \aghethree-$^3$He complex can be written as \cite{krems:ybf:pra2007}
\begin{equation}
\hat{H} = -\frac{\hbar^2}{2\mu R}\frac{\partial^2}{\partial R^2}R +\frac{\bvec{L}^2}{2\mu R^2} + V(R,r,\theta) + \hat{H}_\text{mol},
\end{equation}
where $R$ stands for the atom-molecule separation, $r$ is the internuclear distance in AgHe, $\theta$ is the angle between the unit vectors $\hat{r}=\bvec{r}/r$ and $\hat{R}=\bvec{R}/R$, $\bvec{L}^2$ is the orbital angular momentum for the collision, $\mu$ is the AgHe-He reduced mass, $V(R,r,\theta)$ is the interaction potential, and $\hat{H}_\text{mol}$ is given by \eqn{eqn:HMol}. The eigenstates of $\hat{H}_\text{mol}$ are the Zeeman energy levels of AgHe shown in \fig{fig:agHeZeeman}. We choose the following low-field-seeking states of AgHe as the initial states for scattering calculations: $|N,m_N,m_I,m_S\rangle = |0,0,\sfrac 12,\sfrac 12\rangle$ and $|1,0,\sfrac 12,\sfrac 12\rangle$.

\begin{figure}[t] 
	\centering
		\includegraphics[width=0.95\linewidth]{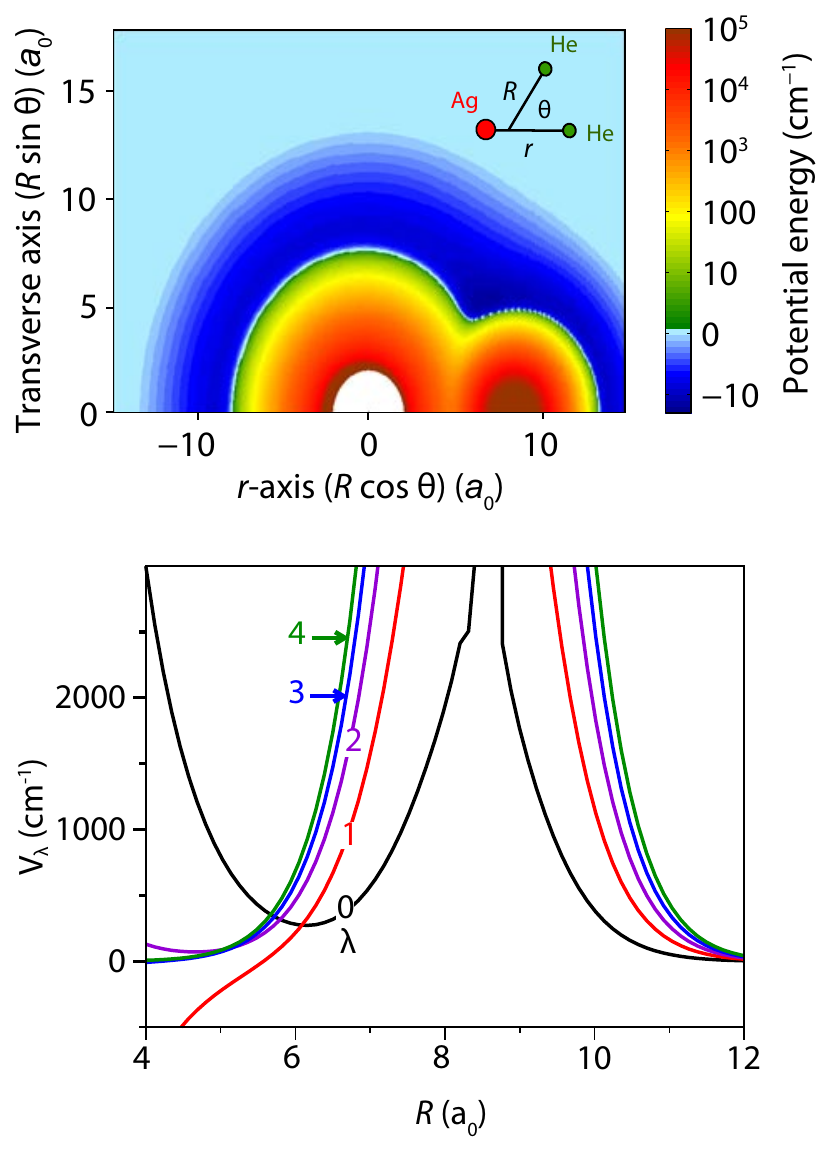}
	\caption{Top: Pairwise AgHe$_2$ interaction PES, evaluated at $r = 8.75~a_0$, corresponding to the bottom of the AgHe potential well.  Color scale is logarithmic, with green to red indicating energies from 1 to $10^5~\mathrm{cm}^{-1}$, and cyan to blue indicating energies from $-1$ to $-15~\mathrm{cm}^{-1}$.  Bottom: Lowest-order expansion coefficients $V_\lambda(R)$ of the AgHe$_2$ PES as functions of $R$, evaluated with $r=3.0~a_0$.  Note the divergence in the vicinity of $R=8.5~a_0$.}
	\label{fig:agHeHePes}
\end{figure}

The \aghethree$_2$ interaction potential is represented as the sum of the pairwise Ag--He and He--He potentials. We express the potential in the Jacobi coordinates illustrated in \fig{fig:agHeHePes}: $R$ -- the distance between the molecule's center of mass and the colliding He atom, $r$ -- the molecule's internuclear separation, and $\theta$ -- the angle between these two vectors.  The He--He potential is taken from Ref.~\citenum{jeziorska:hedimer}.  The number of terms in the Legendre expansion of the interaction potential (\ref{eqn:LegendreExpansion}) increases with increasing $r$, as the interaction potential becomes more anisotropic. At $r>r_c$, where $r_c$ is some critical value, the topology of the interaction potential changes dramatically and the expansion in \eqn{eqn:LegendreExpansion} becomes inadequate, as illustrated in the lower panel of \fig{fig:agHeHePes}. The changes in topology include the appearance of short-range minima corresponding to the insertion of the He atom into the stretched AgHe bond. Furthermore, the pairwise additive approximation is expected to fail at short range, which may lead to unphysical effects in the three-body exchange region. In order to avoid these problems, we choose to fix $r$ at 3.0 $a_0$
rather then keeping the real AgHe equilibrium distance ($r=8.5$ $a_0$). This procedure is consistent with the assumption of negligible direct three-body processes (see \S\ref{ss:aghe:formation}).

The wave function of the \aghethree$_2$ complex is expanded in the fully uncoupled basis \cite{krems:ybf:pra2007}
\begin{equation}\label{eqn:UncoupledBasis}
|Nm_N\rangle |Sm_S\rangle |Im_I\rangle |Lm_L\rangle,
\end{equation}
where $|Lm_L\rangle$ are the partial waves describing the orbital motion of the collision partners. The asymptotic behavior of the expansion coefficients defines the scattering matrix and the probabilities for collision-induced transitions between the different Zeeman states of AgHe. As in the case of Ag-$^3$He collisions described above, the scattering boundary conditions are applied in the basis \cite{krems:ybf:pra2007} which diagonalizes the asymptotic Hamiltonian $\hat{H}_\text{mol}$.
 The asymptotic transformation mixes different
$M_N$ and $M_S$, but is diagonal in $L$ and $M_L$. 
We integrate the coupled-channel equations for the radial coefficients in \eqn{eqn:UncoupledBasis} numerically in a cycle over the total angular momentum projection $M = m_N+m_S+m_I+m_L$ from $R=3$ $a_0$ to $50$ a$_0$ with a step size of 0.04~$a_0$.  The calculations are converged to better than 50\% with respect to the maximum number of rotational states ($N\le 7$) and partial waves ($L\le 7$) included.


The total molecular spin-change rate $\bar{g}_\mathrm{SC}$ is determined by the thermal and trap average of $g_\mathrm{SC}(E,B)$.  We perform the trap average using the averaging distribution in \eqn{eqn:pscavg} and thermally averaged \aghehecol\ inelastic collision rates calculated on a log-spaced grid of 41 points in the range $B=10^{-4}$ to 5 T as described above.  The contribution of molecular spin relaxation to the overall spin-change rate of Ag within the trap is finally given by $\kappa\nhe\bar{g}_\mathrm{SC}$, and is shown in \fig{fig:agHeHeSpinRate}.  As expected from the discussion in \S\ref{ss:gen:traps}, the \aghehecol\ collisional spin-change rate is far too small to explain the observed trap loss.

\begin{figure} 
	\centering
		\includegraphics[width=0.95\linewidth]{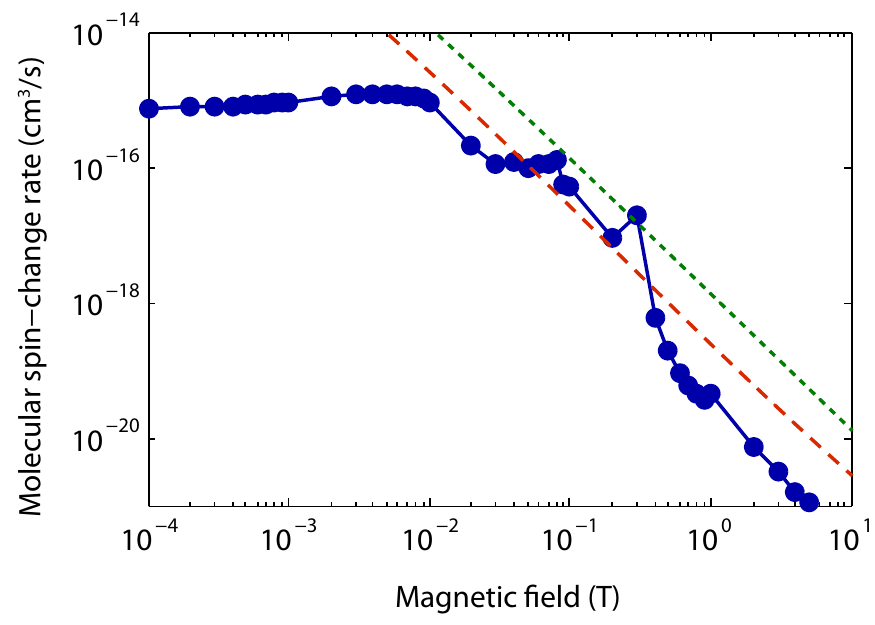}
	\caption{Rate coefficients for \aghehecol\ spin-change as a function of magnetic field, for collision energy $E_\mathrm{col} = 0.5~\textrm{cm}^{-1}$.  Circles are results of the coupled-channel calculations for scaled spin-rotation.  The lines are the asymptotic forms of \eqn{eqn:pscavg}, evaluated for the isotropic (red dashed line) and anisotropic (green dotted line) hyperfine interactions.}
	\label{fig:agHeRateBar}
\end{figure}

As shown in \fig{fig:agHeRBC}, the rigid-rotor approximation can underestimate the three-body recombination rates by as much as factor of 10, and it is not unreasonable to expect a similar level of performance for spin relaxation rates.  A fully quantum theory of molecular spin relaxation in the presence of chemical exchange and three-body breakup channels in a magnetic field will be needed to fully understand the dynamics of formation and spin relaxation of vdW molecules in magnetic traps. 

\subsection{Adiabatic transitions} 
\label{ss:aghe:lz}

Finally, we identify an additional route to spin change in \hethree-containing vdW molecules.  Because of the $r$-dependence and tensor-characteristic of the anisotropic hyperfine interaction, it can mix states of different $N$ quantum numbers.  This occurs when a state with quantum numbers $\ket{N,m_N,m_S,m_{I\ag},m_{I\he}}$ experiences an energy crossing with a state with quantum numbers $\ket{N+2,m_N',m_S',m_{I\ag},m_{I\he}'}$, where the condition $m_N+m_S+m_{I\he}= m_N'+m_S'+m_{I\he}'$ is met. The latter condition follows from the symmetry properties of the matrix elements of the anisotropic hyperfine interaction in the fully uncoupled basis $|Nm_N\rangle |Sm_S\rangle |I_\ag m_{I\ag}\rangle |I_\he m_{I\he}\rangle$ (see, e.g., Eq. 8 of Ref. \citenum{krems:ybf:pra2007}).  For the $\ket{0,0,\sfrac 12,m_{I\he}}$ states of \aghethree, eight crossings, shown in \fig{fig:agHeZeeman}, and tabulated in Table~\ref{table:aghecrossings}, 
occur at magnetic fields of 1.063~T and 1.125~T.

\begin{table}
\caption{Avoided crossings between the $N\mathord=0$ and $N\mathord=2$ rotational levels of \aghethree.  Crossings are between states with $\ket{N,m_N,m_S,m_{I\he}} = \ket{0,0,\sfrac 12,m_{I\he}}$ and $\ket{2,m_N',-\sfrac 12,m_{I\he}'}$.  The Ag nuclear spin projection $m_{I\ag}$ is conserved at the crossing.  At the crossings, occurring at magnetic field $B_\mathrm{LZ}$, energy levels are split by $2 \hbar \Omega$, where $\hbar \Omega$ is the matrix element coupling the two levels.}
\label{table:aghecrossings}
\begin{revtabular}{cccccc}
$m_{I\ag}$	&	$m_{I\he}$ & $m_N'$ & $m_{I\he}'$ & $\Omega/2\pi$~(kHz)	& $B_\mathrm{LZ}$~(T) \\
\midrule
$-\sfrac 12$	&	$\sfrac 12$	&	$-2$	&	$-\sfrac 12$	&	190	&	1.063	\\
					&					&	$-1$	&	$\sfrac 12$		&	95		&			\\
					&	$-\sfrac 12$&	$-1$	&	$-\sfrac 12$	&	95		&			\\
					&					&	$0$	&	$\sfrac 12$		&	39		&			\\
$\sfrac 12$		&	$\sfrac 12$	&	$-2$	&	$-\sfrac 12$	&	190	&	1.125	\\
					&					&	$-1$	&	$\sfrac 12$		&	95 	&			\\
					&	$-\sfrac 12$&	$-1$	&	$-\sfrac 12$	&	95 	&			\\
					&					&	$0$	&	$\sfrac 12$		&	39		&			
\end{revtabular}
\end{table}

Consider a trapped \aghethree\ molecule orbiting within the trap.  As the molecule crosses the spatial shell where the magnetic field causes a crossing, it has a chance of transiting adiabatically, thus resulting in a spin flip.  The probability that the molecule follows adiabatically is given by the Landau-Zener formula:
\begin{equation}
p_\textrm{LZ}=1-\exp \left ( -\frac{\hbar \pi \Omega^2}{\mu_B\, \bvec{v}\cdot\bvec{\nabla} B} \right ),
\label{eqn:lzprob}
\end{equation}
where $\hbar\Omega$ is the matrix element coupling the trapped and untrapped states.

In the limit that the fraction of molecules flipped per unit time is small, and in the limit that flipped atoms are ejected rapidly, without opportunity to cross the Landau-Zener region a second time, we may estimate the trap loss rate from the flux of trapped molecules across the Landau-Zener region:
\begin{multline}
\dot{N}_\mathrm{LZ} \sim \sum_{I_\mathrm{He},I_\mathrm{Ag}}
\pi u_{\textrm{LZ}}^2 n(u_\textrm{LZ}) \\
\times \frac { \int\!\int g(\bvec v)\, p_\textrm{LZ} \! \left (\bvec v\mathord\cdot\bvec \nabla B\right ) \bvec v\mathord\cdot\bvec{\hat{u}}\,\cos\theta d\theta d\bvec v }
{\int g(\bvec v) d\bvec v}.
\label{eqn:lzapprox}
\end{multline}
Here the local density $n(u)$ of trapped molecules is a function of $\bvec u\equiv\bvec r + 2 \bvec z$.  $u_\textrm{LZ}$ is the value of $u$ at the Landau-Zener region, with $u_\textrm{LZ}$ given by the ratio of the crossing field $B_\mathrm{LZ}$ and the field gradient $B'$, and $\theta = \arctan \left ( r/2z \right )$.  $g(\bvec v) = \exp(- m v^2/2 k_B T)$ is the Boltzmann factor.  For high thermal velocity, and assuming that the distribution of trapped molecules is in thermal equilibrium throughout the trap, we find that the effective spin-loss rate is
\begin{equation}
k_\textrm{LZ} = \frac{\dot{N}_\mathrm{LZ}}{n_\mathrm{He}} \sim \sum_{I_\mathrm{He},I_\mathrm{Ag}} \frac{1}{4} \frac{h \Omega^2 u_\textrm{LZ}^2}{\mu_B\,B'} \frac{e^{-B' u_\textrm{LZ}/k_B T}}{V_\mathrm{eff}} \kappa,
\label{eqn:lzanalytic}
\end{equation}
where $V_\mathrm{eff} = 4\pi (k_B T/\mu_B B')^3$ is the effective volume of the trap.

In the above calculation, we made many approximations which may not be satisfied in the actual system.  In order to calculate the adiabatic transition loss rate when these approximations do not hold, we used a semiclassical direct-simulation Monte Carlo approach to calculate the system dynamics.  We initialize the calculation by generating a sample of Boltzmann-distributed free Ag atoms.  The atoms evolve under the trap magnetic field.  After a random time, exponentially distributed with mean value equal to the mean free atom--He collision rate, we (classically) simulate an elastic collision with a He atom randomly generated from a Boltzmann distribution.  This process of evolution and collisions continues until the atom leaves the magnetic trap.  Each collision has a chance of causing molecular formation (in a random $^3$He Zeeman state), with mean formation time equal to $1/K \nhe^2$.  Similarly, collisions of bound \aghethree\ with free \hethree\ have a random chance to cause dissociation, with mean dissociation time equal to $1/D \nhe$.   We neglect the energy released and absorbed in formation and dissociation, and we neglect the rotational relaxation dynamics, as both of these degrees of freedom equilibrate quickly.  When a bound molecule crosses the Landau-Zener region, it spin flips with a probability calculated using \eqn{eqn:lzprob}.  Spin flips can occur both from the trapped to the untrapped state and \textit{vice versa}.  When a bound or free atom reaches the edge of the trap, it is removed from the simulation.  For each value of $T$, we simulate six to eight values of $\nhe$ using 4000 atoms for each simulation.  We extract the atomic lifetime $\tau$ from the simulation, then fit $\tau$ vs.\ $\nhe$ using Eqn.~6 of Ref.~\citenum{brahms:agTrapping}, giving the spin-change rate coefficient as a function of temperature.

We find that this calculation agrees with the approximate analytic expression of \eqn{eqn:lzanalytic} to within a factor of four for the experimental parameters.  Furthermore, the result reproduces the experimentally observed temperature dependence.  The overall magnitude of the calculation underpredicts the experimental data, but this can be explained by increasing the binding energy to $1.53~\mathrm{cm}^{-1}$, a 10\% larger value than we predict.  However, we note that the AgHe potential of Cargnoni \textit{et al.} \cite{cargnoni:mHePotentials} also predicts a binding energy of $1.53~\mathrm{cm}^{-1}$.  This adjusted result is compared to data in Fig.~1 of Ref.\ \citenum{brahms:vdw}.  Based on our analysis of Landau-Zener-induced spin flips, we conclude that these adiabatic transitions are responsible for the experimentally observed trap loss.

\section{Spectroscopy} 
\label{ss:spectro}

We now turn to the possibility of spectroscopic detection of vdW molecules.  In general, we might expect that the molecular spectrum is similar to the bare atomic spectrum of Ag, due to the rather weak energy shifts of the molecular bound states from the continuum states.

One major impediment to spectroscopic detection is the possibility of photodissociation.  This can occur through two possible channels.  First, because the molecules are weakly bound, only a few vibrational levels typically exist in the ground manifold.  The overlap between the excited electronic vibrational states and the nuclear continuum may therefore be significant, and photodissociation can occur either by direct excitation to the continuum, or during spontaneous decay back to the ground state \cite{levy:science:vdw}.  A third mechanism, predissociation \cite{miller:vdw:review}, is present in molecules for which the potential energy surface (PES) of the bound excited electronic state intersects the PES of a continuum excited state.  In \aghe, this intersection occurs between the ${^2}\Pi_{3/2}$ PES and the manifold of surfaces that asymptotically approach the Ag ${^2}D$ state (see \fig{fig:agHePotentials}).  In such a system, the molecule can dissociate via this coupling.

When vdW molecules are produced in very dense He environments (e.g., He nanodroplets), the collisional formation and dissociation rates can be much higher than the photodissociation rate, and the equilibrium population of vdWs will in general remain high.  In buffer-gas cooling experiments, however, the photodissociation rate can easily exceed the collisional rate for return to chemical equilibrium.  

We model photodissociation by assuming that dissociation occurs at a rate $\Gamma_\mathrm{dis}$ proportional to the rate $\Gamma_\mathrm{abs}$ of photon absorption.  For isolated transitions, where quantum interference can be neglected,
\begin{equation}
\Gamma_\mathrm{dis} = p_\mathrm{dis} \Gamma_\mathrm{abs},
\end{equation}
where the branching ratio $p_\mathrm{dis}$ is given by the ratio of the decay rate to unbound states $\gunbound$ to the total decay rate $\gtotal$.  The photoabsorption rate per molecule is \cite{demtroder}
\begin{equation}
\Gamma_\mathrm{abs} = \frac{I \sigma_\gamma \lambda}{h c} \frac{\gtotal^2/4}{\delta^2+\gtotal^2/4}, 
\end{equation}
where $c$ is the speed of light, $I$ is the intensity of the pump beam, $\lambda$ is the pump wavelength, and $\sigma_\gamma$ is the on-resonance photon absorption cross-section.  The detuning of the pump from molecular resonance is $\delta$.  The equilibrium population of molecules in rovibrational state $i$ in the presence of a photodissociating spectroscopy beam can be calculated from detailed balance:
\begin{equation}
n_i(I,\delta) =  \frac{\kappa_i(T)\nx\nhe}{1 + \Gamma_\mathrm{dis}/D_i(T)\nhe }.
\end{equation}
The above equation indicates that the molecule population will be depleted when the light intensity exceeds a saturation intensity defined by
\begin{equation}
I_{s;i} = \frac{h c D_i \nhe}{\sigma_\gamma \lambda p_{\mathrm{dis};i}}.
\end{equation}
For a molecular state with $\sigma_\gamma = 3\lambda^2/2\pi$, $D \nhe = 100~\mathrm{s}^{-1}$, $p_\mathrm{dis} = 0.01$, and $\lambda = 300~\mathrm{nm}$, the saturation itensity is approximately 100~nW/mm$^2$.

Using this general model, we now estimate the applicability of various spectroscopy techniques, including absorption spectroscopy, laser-induced fluorescence, and ionization.  We also propose a spectroscopy method which relies on arresting the overall atomic spin change by preventing molecular formation.

\subsubsection{Absorption.~~}  The optical density of a molecular vapor is
\begin{equation}
-\ln \left ( I/I_0 \right ) = \int \sigma_\gamma \nxhe(z) \frac{\gtotal^2/4}{\delta^2+\gtotal^2/4} dz.
\end{equation}
In the limit of weak absorption (optical density $\ll 1$), such that the light intensity is approximately constant throughout the vapor, the optical density becomes
\begin{equation}
I/I_0 = \int \sigma_\gamma \nx(z) \kappa \nhe \frac{\gtotal^2/4}{\delta^2 + (1+I_0/I_s) \gtotal^2/4} dz,
\end{equation}
and decreases as $\sim 1/I_0$ for $I_0 > I_s$.  This indicates that rather weak probe beams are necessary to avoid dissociation-induced power-broadening of the molecular line.

\subsubsection{Laser-induced fluorescence.~~}  Due to the saturation effect discussed above, the maximum rate of photon scatter will be limited by photodissociation.  The photon scattering rate is
\begin{align}
\Gamma_\mathrm{scat} &= \frac{I_0 \sigma_\gamma \lambda}{h c} \kappa \nhe \frac{\gtotal^2/4}{\delta^2 + (1+I_0/I_s) \gtotal^2/4} \int \nx \dr. \\
&\sim \frac{K \nhe^2}{p_\mathrm{dis}} \int \nx\dr \quad\quad \textrm{for }{I_0 \gg I_s}.
\end{align}
The second expression holds for strong probe beams, showing that the maximum scattering rate is proportional to the recombination rate.  For $K = 10^{-31}~\mathrm{cm^6/s}$, $\nhe = 3\times 10^{16}~\cmthree$, and $p_\mathrm{dis} = 0.01$, the rate is approximately $10^4~\mathrm{s^{-1}}$ per $X$ atom.

\subsubsection{Ionization.~~}  Because of the difficulty of collecting large numbers of photons for each $X$He molecule, ionization methods may be a sensitive probe for positive detection of van der Waals molecules formed in these experiments.  In particular, resonantly-enhanced multi-photon ionization should also give information on the molecular structure, subject to the power- and dissociation-broadening effects described above.  Care must also be taken to keep electric fields small enough to prevent DC discharge of the He gas (potential $\lesssim 20$~V).

\subsubsection{Spin-change spectroscopy.~~} For systems such as \aghethree, in which photon absorption causes rapid molecular dissociation, and for which Landau-Zener transitions cause rapid spin relaxation, an alternate method of spectroscopy should become available.  By applying a beam with intensity of a few $I_s$ throughout the gas, tuned to a spin-preserving ``stretched'' transition, it should be possible to depopulate the molecular population, thereby preventing molecule-mediated spin-change.  The spectroscopic signal in this case would be extracted from the rate of spin change vs. the detuning of the dissociating beam.  

\subsection{AgHe spectroscopy}

\subsubsection{Theory.~~}
In this section, we evaluate the probabilities for electric dipole transitions between the ground $^2\Sigma_{1/2}$ and excited (${^2}\Pi_{3/2}$) electronic states of AgHe. 
As shown in the Electronic Information, the transition probability of the AgHe molecule relative to the free Ag atom is given by
\begin{multline}\label{eqn:spectro:Msummed32}
P_{rel}(nvJ \to v'J'J_a'\Omega'=\textstyle\frac{3}{2} ) \propto (2J'+1)(2J+1) \\
\times \langle \chi_{v'J'J_a'\Omega'}(r)| \chi_{vJ}(r)\rangle ^2 
\threejm{J'}{\Omega'}{1}{\Omega-\Omega'}{J}{-\Omega}^2,
\end{multline}
where $v$ is the vibrational quantum number, $J_a$ is the total electronic angular momentum of Ag (approximately conserved in the molecule), $J = N+J_a$ is the total angular momentum of Ag, and $\Omega$ is the projection of $J$ on the internuclear axis. The primes in Eq. (\ref{eqn:spectro:Msummed32}) refer to the quantum numbers of the excited $^2\Pi_{3/2}$ state (see Fig.\ref{fig:agHePotentials}). The Franck-Condon overlaps  $f_\mathrm{FC} = \langle\chi_{v'J'J_a'\Omega'}(r)| \chi_{vJ}(r)\rangle^2$ were constructed by numerical integration using the {\it ab initio} potential energy curves for the ground and excited electronic states of AgHe calculated in Sec. \ref{sec:MolecularStructure}. 

\begin{figure} 
	\centering
		\includegraphics[width=0.95\linewidth]{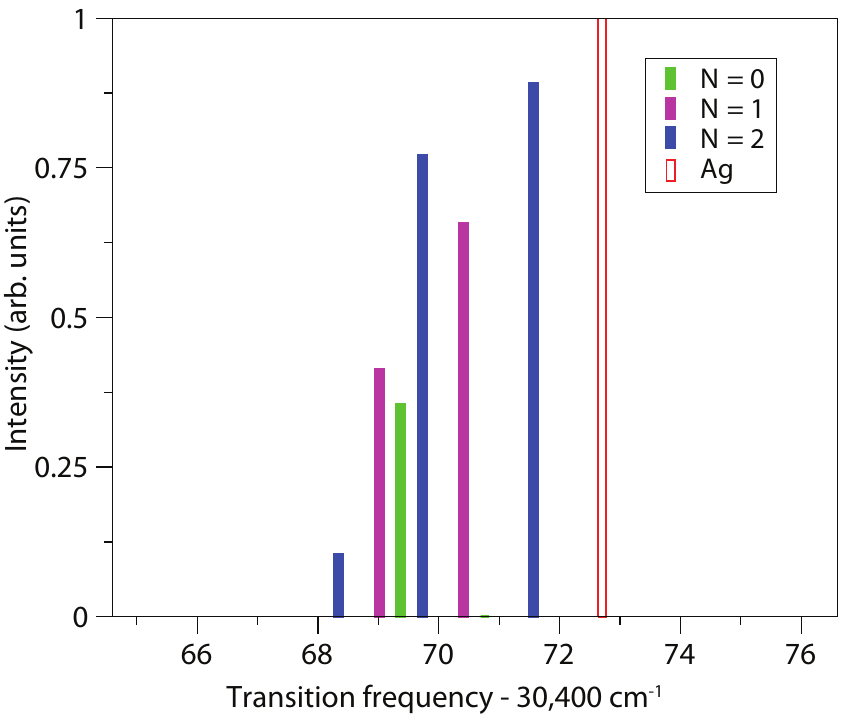}
	\caption{Theoretical stick spectrum of the $^2\Pi_{3/2} \leftarrow ^2\Sigma_{1/2}$  transition in AgHe. The transition energy is given in units of $\Delta-30,400$ cm$^{-1}$, where $\Delta$ is the transition frequency. The red line marks the position of the Ag $D_2$ line.}
	\label{fig:agHeSpectrum}
\end{figure}

\fig{fig:agHeSpectrum} shows the calculated stick spectrum for the \mbox{$^2\Sigma \to {^2\Pi_{3/2}}$} transition in AgHe in the vicinity of the $D_2$ atomic transition in Ag. Transition energies and Franck-Condon factors for individual vibrational levels are listed in Table~\ref{table:FCF}. In the following, we focus exclusively on the most efficient electronic transition $^2\Sigma \to {^2}\Pi_{3/2}$; the probability for the transition $^2\Sigma \to {^2}\Pi_{1/2}$ is very small because of the negligible Franck-Condon overlap between the $^2\Sigma$ and $^2\Pi_{1/2}$ electronic states (see Fig. \ref{fig:agHePotentials}).

The spectrum contains a number of transitions from different initial rotational levels of the $v=0$ vibrational level. For convenience, the states are labeled with their Hund's case (b) quantum number $N$. We neglect the weak spin-rotation interaction in the ground $^2\Sigma$ state, so the $N\pm 1/2$ components are degenerate and only the transitions from the $J=N+1/2$ component of each $N$-state are displayed in Fig \ref{fig:agHeSpectrum}. 

From \fig{fig:agHeSpectrum} and Table \ref{table:FCF}, we observe that transitions from the ground $^2\Sigma,v=0,J'=0$ state occur predominantly to the most weakly bound $v'=4$ level of the $^2\Pi_{3/2}$ electronic state. Transitions to the next most deeply bound $v'=3$ level are suppressed by a factor of $\sim$10 due to the diminishing Franck-Condon overlap with the ground state, and transitions to the $v'\le2$ levels have negligible probabilities. Thus, our calculations suggest that only the highest $v'=3,4$ vibrational levels supported by the $^2\Pi_{3/2}$ electronic state can be populated in either absorption or fluorescence spectroscopy. As shown in Fig. \ref{fig:agHePotentials}, these vibrational levels lie  above the avoided crossing of the $^2\Pi_{3/2}$ electronic states correlating with the $^2P_{3/2}$ and $^2D_{5/2}$ dissociation limits. The crossing occurs $\sim$115 cm$^{-1}$ below the dissociation limit of the $^2\Pi_{3/2}$ state, leading to the possibility of electronic predissociation via non-adiabatic transitions. The predissociation will shorten the lifetime of the $v'=3,4$ states, but is unlikely to make their observation impossible, since Jouvet {\it et al.} were able to observe highly excited vibrational levels of the AgAr complex  \cite{jouvet:AgAr} via laser fluorescence excitation spectroscopy, even though the non-adiabatic couplings in AgAr are much stronger than in AgHe. 

The rotational structure of each vibrational band is determined by the $\Delta J=0,\pm 1$ selection rule, so there is only one transition from $N=0$, two transitions from $N=1$, and three transitions from $N=2$. The relative intensities of different rotational transitions shown in \fig{fig:agHeSpectrum} are set by thermal populations of the $N=0-2$ rotational levels of the ground electronic state, which are determined by the rotational temperature of AgHe. The line intensities, relative to the intensity of the atomic $D_2$ transition, are equal to $\kappa(T) n_\text{He} f_\mathrm{FC}$.

\begin{table*}
\centering
\caption{Rovibrational states involved in transitions shown in Fig. 2. The energies of the states (in cm$^{-1}$) are shown
relative to their own dissociation limits. The transition frequencies and plotted in Fig.~2. The binding energies are calculated using the {\it ab initio}  interaction potentials for the excited electronic states of AgHe computed in the present work (see Sec. \ref{sec:MolecularStructure}) and the ground-state AgHe potential from Ref.~\citenum{cargnoni:mHePotentials}. The level $|v'=4,J'=9/2\rangle$ is unbound. The Franck-Condon overlaps between the initial $\langle^2X,v=2,J|$ and final $|^2\Pi_{3/2},v'<4,J'\rangle$ levels are smaller than 0.1 for all $J'$. }

\label{table:FCF}
\vspace{0.3cm}
\begin{revtabular}{cccccc}
Initial state &  Binding  &  Final state     & Binding& $\Delta$            & Franck-Condon \\
$|v,N,J\rangle$ & energy  &  $|v',J'\rangle$ & energy & $-30,400$~cm$^{-1}$ & factor \\
\midrule
$|0,0,\sfrac 12\rangle$  &    $-1.5279$           &  $|4,\sfrac 32\rangle$   & $-4.8581$   &  69.373 & 0.709 \\
$|0,1,\sfrac 32\rangle$  &    $-1.1715$           &  $|4,\sfrac 32\rangle$   & $-4.8581$   &  69.016 & 0.689 \\
                  &                             &  $|4,\sfrac 52\rangle$   & $-3.4691$   &  70.405 & 0.730 \\
$|0,2,\sfrac 52\rangle$  &    $-0.4914$           &  $|4,\sfrac 32\rangle$   & $-4.8581$   &  68.336 & 0.629 \\
                  &                             &  $|4,\sfrac 52\rangle$   & $-3.4691$   &  69.725 & 0.675 \\
                  &                             &  $|4,\sfrac 72\rangle$   & $-1.6305$   &  71.564 & 0.749
\end{revtabular}
\end{table*}

\subsubsection{Experiment.~~}
An attempt was made to spectroscopically observe \aghethree\ molecules, using the apparatus of Ref.~\citenum{brahms:agTrapping}.  Only absorption spectroscopy is possible in this apparatus.
A frequency-doubled dye laser (Coherent 899) operating at 328~nm, having $\approx 1$~MHz linewidth, was used to produce a probe beam with 500~nW/mm$^2$ intensity.  Approximately $5\times 10^{10}$ Ag were ablated into a 330~mK \hethree\ buffer gas with $\nhe \approx 3\times 10^{16}~\cmthree$, yielding an optical absorption on the atomic $D_2$ line of $e^{-2}$.  At these parameters, assuming Franck-Condon factors for the molecular transition on the order of $0.9$, we therefore expect absorption of $e^{-0.3}$.  However, no absorption was detected, with an absorption sensitivity of $e^{-0.003}$, with the laser scanned from $30,438.0$ to $30,476.0~\cmone$.  We therefore conclude that either: the \aghethree\ formation rate is no larger than $10^{-35}~\mathrm{cm^6/s}$, such that the system did not enter thermal equilibrium on the experimental diffusion time scale, or that the photodissociation probability per absorbed photon is near unity, such that the \aghethree\ population was rapidly depleted below the experimental detection sensitivity.

\section{Conclusion}
\label{ss:conclusion}

We have described how a wide variety of He-containing vdW complexes can be formed in buffer-gas cooling experiments.  In contrast to formation in He nanodroplets, the molecules formed here exist in a dilute environment.  We have shown how the spin stability of species in buffer-gas loaded magnetic traps can be uniquely sensitive to the formation and dynamics of vdW molecules.  With AgHe, this sensitivity allows the observation of novel trap spin dynamics, mediated by the anisotropic hyperfine interaction.  We have also detailed spectroscopy in this system, showing that care must be taken to avoid photodissociation of the vdW molecules in traditional spectroscopy, while sensitivity to the vdWs molecules' magnetic moments allows for a novel spectroscopic method.

It may also be possible to trap these complexes by rapidly removing the He buffer gas from the trap.  Such a removal process has been demonstrated both with trapped alkali and transition metal atoms, leaving dense samples trapped for hundreds of seconds\cite{brahms:thesis}.  Buffer gas can be removed on timescales $\tau_r$ smaller than tens of milliseconds.  During buffer gas removal, the dissociation of vdW molecules will cease when the buffer gas density falls below $1/D \tau_r$.  In the limit of low buffer-gas density and low vdW molecule density\footnote{We assume the temperature is low enough that Landau-Zener loss does not occur, or that the molecule is composed using $^4$He.}, the molecular trap lifetime will be limited instead by (1) $X$He--$X$ dissociating collisions, (2) $X$He--$X$ spin-changing collisions, and (3) three-body collisions.  These rates we expect to be small, due to (1) low collision energy, (2) low probability of spin-change in $X$--$X$ collisions, and (3) low $X$ and $X$He density.  Nuclear exchange collisions \mbox{($X\he + X \leftrightarrow X_2 + \he$)} may cause limited lifetimes.  For systems where the nuclear exchange rate is small, it may be possible to sympathetically cool the trapped gas with evaporatively cooled trapped $X$, allowing cooling of the trapped vdW molecules to the ultracold regime.  For systems where the exchange rate is high, the $X$ can be optically removed from the trap, leaving a long-lived sample of molecules.

 As mentioned in the Introduction, vdW complexes (and molecular clusters in general) provide ideal prototype systems for describing a wide variety of phenomena in chemistry and physics. The creation of cold, trapped vdW complexes may thus open new avenues of research in few-body physics, chemical reaction dynamics, and cluster physics. In particular, it would be interesting to explore the possibility of controlling three-body recombination with external electromagnetic fields, which would enable the production of size-selected vdW clusters, and may allow the study of exotic few-body phenomena such as the Efimov effect \cite{stoll:he3,villareal:he3,esry:He3}.   Previous experimental \cite{herschbach:Rg2} and theoretical  \cite{gettys:JPC88:QCT,kalus:JCP99:QCT}  studies of chemical reactions between vdW molecules have been limited to high collision energies.  Table \ref{table:dimerEnergies} illustrates that most He-containing  vdW molecules have binding energies that are comparable with Zeeman shifts induced by magnetic fields easily achievable in the laboratory. Thus it is natural to expect that chemical reactions of cold vdW molecules should be particularly amenable to external field control. Further experimental and theoretical studies of these and related phenomena (such as collision-induced dissociation and predissociation) will greatly enhance our ability to understand and control intermolecular interactions and few-body phenomena at low temperatures.

\section*{Acknowledgements}
We thank Alexei Buchachenko for calculating the binding energies of YbF-He and CaH-He.  This work was supported by NSF grants to the Harvard-MIT Center for Ultracold Atoms and the Institute for Theoretical Atomic, Molecular and Optical Physics at Harvard University and the Smithsonian Astrophysical Observatory.  T.W.'s work was supported by the NSF and the DOE Office of Basic Energy Sciences, grant \#DE-FG02-03ER46093.

\clearpage

\providecommand*{\mcitethebibliography}{\thebibliography}
\csname @ifundefined\endcsname{endmcitethebibliography}
{\let\endmcitethebibliography\endthebibliography}{}


\begin{mcitethebibliography}{93}
\providecommand*{\natexlab}[1]{#1}
\providecommand*{\mciteSetBstSublistMode}[1]{}
\providecommand*{\mciteSetBstMaxWidthForm}[2]{}
\providecommand*{\mciteBstWouldAddEndPuncttrue}
  {\def\EndOfBibitem{\unskip.}}
\providecommand*{\mciteBstWouldAddEndPunctfalse}
  {\let\EndOfBibitem\relax}
\providecommand*{\mciteSetBstMidEndSepPunct}[3]{}
\providecommand*{\mciteSetBstSublistLabelBeginEnd}[3]{}
\providecommand*{\EndOfBibitem}{}
\mciteSetBstSublistMode{f}
\mciteSetBstMaxWidthForm{subitem}
{(\emph{\alph{mcitesubitemcount}})}
\mciteSetBstSublistLabelBeginEnd{\mcitemaxwidthsubitemform\space}
{\relax}{\relax}

\bibitem[Carr \emph{et~al.}(2009)Carr, DeMille, Krems, and
  Ye]{carr:NJP:review2009}
L.~D. Carr, D.~DeMille, R.~V. Krems and J.~Ye, \emph{New J. Phys.}, 2009,
  \textbf{11}, 055049\relax
\mciteBstWouldAddEndPuncttrue
\mciteSetBstMidEndSepPunct{\mcitedefaultmidpunct}
{\mcitedefaultendpunct}{\mcitedefaultseppunct}\relax
\EndOfBibitem
\bibitem[Bell and Softley(2009)]{softley:molphys:review}
M.~T. Bell and T.~P. Softley, \emph{Mol. Phys.}, 2009, \textbf{107}, 99\relax
\mciteBstWouldAddEndPuncttrue
\mciteSetBstMidEndSepPunct{\mcitedefaultmidpunct}
{\mcitedefaultendpunct}{\mcitedefaultseppunct}\relax
\EndOfBibitem
\bibitem[Schnell and Meijer(2009)]{schnell:angewchem:review}
M.~Schnell and G.~Meijer, \emph{Angew. Chem. Int. Ed.}, 2009, \textbf{48},
  6010\relax
\mciteBstWouldAddEndPuncttrue
\mciteSetBstMidEndSepPunct{\mcitedefaultmidpunct}
{\mcitedefaultendpunct}{\mcitedefaultseppunct}\relax
\EndOfBibitem
\bibitem[Krems \emph{et~al.}(2009)Krems, Stwalley, and
  Friedrich]{ColdMoleculesBook}
\emph{Cold Molecules: Theory, Experiment, Applications}, ed. R.~V. Krems, W.~C.
  Stwalley and B.~Friedrich, CRC Press, New York, 2009\relax
\mciteBstWouldAddEndPuncttrue
\mciteSetBstMidEndSepPunct{\mcitedefaultmidpunct}
{\mcitedefaultendpunct}{\mcitedefaultseppunct}\relax
\EndOfBibitem
\bibitem[Meyer \emph{et~al.}(2006)Meyer, Bohn, and Deskevich]{bohn:pra:edm}
E.~R. Meyer, J.~L. Bohn and M.~P. Deskevich, \emph{Phys. Rev. A}, 2006,
  \textbf{73}, 062108\relax
\mciteBstWouldAddEndPuncttrue
\mciteSetBstMidEndSepPunct{\mcitedefaultmidpunct}
{\mcitedefaultendpunct}{\mcitedefaultseppunct}\relax
\EndOfBibitem
\bibitem[Hudson \emph{et~al.}(2002)Hudson, Sauer, Tarbutt, and
  Hinds]{hudson:prl:edm}
J.~J. Hudson, B.~E. Sauer, M.~R. Tarbutt and E.~A. Hinds, \emph{Phys. Rev.
  Lett.}, 2002, \textbf{89}, 023003\relax
\mciteBstWouldAddEndPuncttrue
\mciteSetBstMidEndSepPunct{\mcitedefaultmidpunct}
{\mcitedefaultendpunct}{\mcitedefaultseppunct}\relax
\EndOfBibitem
\bibitem[DeMille(2002)]{demille:prl:polarqi}
D.~DeMille, \emph{Phys. Rev. Lett.}, 2002, \textbf{88}, 067901\relax
\mciteBstWouldAddEndPuncttrue
\mciteSetBstMidEndSepPunct{\mcitedefaultmidpunct}
{\mcitedefaultendpunct}{\mcitedefaultseppunct}\relax
\EndOfBibitem
\bibitem[Krems(2008)]{krems:pccp:coldchem}
R.~V. Krems, \emph{Phys. Chem. Chem. Phys.}, 2008, \textbf{10},
  4079--4092\relax
\mciteBstWouldAddEndPuncttrue
\mciteSetBstMidEndSepPunct{\mcitedefaultmidpunct}
{\mcitedefaultendpunct}{\mcitedefaultseppunct}\relax
\EndOfBibitem
\bibitem[Sawyer \emph{et~al.}(2010)Sawyer, Stuhl, Yeo, Tscherbul, Hummon, Xia,
  Klos, Patterson, Doyle, and Ye]{sawyer:arxiv:collision}
B.~C. Sawyer, B.~K. Stuhl, M.~Yeo, T.~V. Tscherbul, M.~T. Hummon, Y.~Xia,
  J.~Klos, D.~Patterson, J.~M. Doyle and J.~Ye, 2010, arXiV:1008.5127\relax
\mciteBstWouldAddEndPuncttrue
\mciteSetBstMidEndSepPunct{\mcitedefaultmidpunct}
{\mcitedefaultendpunct}{\mcitedefaultseppunct}\relax
\EndOfBibitem
\bibitem[Ospelkaus \emph{et~al.}(2010)Ospelkaus, Ni, Wang, de~Miranda,
  Neyenhuis, {Qu\'em\'ener}, Julienne, Bohn, Jin, and
  Ye]{ospelkaus:science:controlledchem}
S.~Ospelkaus, K.-K. Ni, D.~Wang, M.~H.~G. de~Miranda, B.~Neyenhuis,
  G.~{Qu\'em\'ener}, P.~S. Julienne, J.~L. Bohn, D.~S. Jin and J.~Ye,
  \emph{Science}, 2010, \textbf{327}, 853--857\relax
\mciteBstWouldAddEndPuncttrue
\mciteSetBstMidEndSepPunct{\mcitedefaultmidpunct}
{\mcitedefaultendpunct}{\mcitedefaultseppunct}\relax
\EndOfBibitem
\bibitem[Andre \emph{et~al.}(2006)Andre, DeMille, Doyle, Lukin, Maxwell, Rabl,
  Schoelkopf, , and Zoller]{andre:QS}
A.~Andre, D.~DeMille, J.~M. Doyle, M.~D. Lukin, S.~E. Maxwell, P.~Rabl, R.~J.
  Schoelkopf,  and P.~Zoller, \emph{Nature Phys.}, 2006, \textbf{2}, 636\relax
\mciteBstWouldAddEndPuncttrue
\mciteSetBstMidEndSepPunct{\mcitedefaultmidpunct}
{\mcitedefaultendpunct}{\mcitedefaultseppunct}\relax
\EndOfBibitem
\bibitem[Balakrishnan(2004)]{balakrishnan}
N.~Balakrishnan, \emph{J. Chem. Phys.}, 2004, \textbf{121}, 5563--5566\relax
\mciteBstWouldAddEndPuncttrue
\mciteSetBstMidEndSepPunct{\mcitedefaultmidpunct}
{\mcitedefaultendpunct}{\mcitedefaultseppunct}\relax
\EndOfBibitem
\bibitem[Worsnop \emph{et~al.}(1986)Worsnop, Buelow, and
  Herschbach]{herschbach:Rg2}
D.~R. Worsnop, S.~J. Buelow and D.~R. Herschbach, \emph{J. Phys. Chem.}, 1986,
  \textbf{90}, 5121\relax
\mciteBstWouldAddEndPuncttrue
\mciteSetBstMidEndSepPunct{\mcitedefaultmidpunct}
{\mcitedefaultendpunct}{\mcitedefaultseppunct}\relax
\EndOfBibitem
\bibitem[Smalley \emph{et~al.}(1977)Smalley, Auerbach, Fitch, Levy, and
  Wharton]{levy:NaAr}
R.~E. Smalley, D.~A. Auerbach, P.~S.~H. Fitch, D.~H. Levy and L.~Wharton,
  \emph{J. Chem. Phys.}, 1977, \textbf{66}, 3778\relax
\mciteBstWouldAddEndPuncttrue
\mciteSetBstMidEndSepPunct{\mcitedefaultmidpunct}
{\mcitedefaultendpunct}{\mcitedefaultseppunct}\relax
\EndOfBibitem
\bibitem[Kalus(1999)]{kalus:JCP99:QCT}
R.~Kalus, \emph{J. Chem. Phys.}, 1999, \textbf{110}, 3856\relax
\mciteBstWouldAddEndPuncttrue
\mciteSetBstMidEndSepPunct{\mcitedefaultmidpunct}
{\mcitedefaultendpunct}{\mcitedefaultseppunct}\relax
\EndOfBibitem
\bibitem[Gettys \emph{et~al.}(1988)Gettys, Raff, and
  Thompson]{gettys:JPC88:QCT}
N.~S. Gettys, L.~M. Raff and D.~L. Thompson, \emph{J. Phys. Chem.}, 1988,
  \textbf{92}, 5270\relax
\mciteBstWouldAddEndPuncttrue
\mciteSetBstMidEndSepPunct{\mcitedefaultmidpunct}
{\mcitedefaultendpunct}{\mcitedefaultseppunct}\relax
\EndOfBibitem
\bibitem[Hutson(1990)]{hutson:vdw:intermolecularForces:review}
J.~M. Hutson, \emph{Annu. Rev. Phys. Chem.}, 1990, \textbf{41}, 123--154\relax
\mciteBstWouldAddEndPuncttrue
\mciteSetBstMidEndSepPunct{\mcitedefaultmidpunct}
{\mcitedefaultendpunct}{\mcitedefaultseppunct}\relax
\EndOfBibitem
\bibitem[Gong \emph{et~al.}(2008)Gong, Jau, and Happer]{happer:prl:MRg}
F.~Gong, Y.-Y. Jau and W.~Happer, \emph{Phys. Rev. Lett.}, 2008, \textbf{100},
  233002\relax
\mciteBstWouldAddEndPuncttrue
\mciteSetBstMidEndSepPunct{\mcitedefaultmidpunct}
{\mcitedefaultendpunct}{\mcitedefaultseppunct}\relax
\EndOfBibitem
\bibitem[Parker \emph{et~al.}(2002)Parker, Walker, Kendrick, and
  Pack]{parker:CID:jcp}
G.~A. Parker, R.~B. Walker, B.~K. Kendrick and R.~T. Pack, \emph{J. Chem.
  Phys.}, 2002, \textbf{117}, 6083\relax
\mciteBstWouldAddEndPuncttrue
\mciteSetBstMidEndSepPunct{\mcitedefaultmidpunct}
{\mcitedefaultendpunct}{\mcitedefaultseppunct}\relax
\EndOfBibitem
\bibitem[Krems(2004)]{krems:vdw:magneticBreaking}
R.~V. Krems, \emph{Phys. Rev. Lett.}, 2004, \textbf{93}, 013201\relax
\mciteBstWouldAddEndPuncttrue
\mciteSetBstMidEndSepPunct{\mcitedefaultmidpunct}
{\mcitedefaultendpunct}{\mcitedefaultseppunct}\relax
\EndOfBibitem
\bibitem[Levy(1981)]{levy:science:vdw}
D.~H. Levy, \emph{Science}, 1981, \textbf{214}, 263\relax
\mciteBstWouldAddEndPuncttrue
\mciteSetBstMidEndSepPunct{\mcitedefaultmidpunct}
{\mcitedefaultendpunct}{\mcitedefaultseppunct}\relax
\EndOfBibitem
\bibitem[Mattison \emph{et~al.}(1974)Mattison, Pritchard, and
  Kleppner]{mattison:KAr}
E.~M. Mattison, D.~E. Pritchard and D.~Kleppner, \emph{Phys. Rev. Lett.}, 1974,
  \textbf{32}, 507--509\relax
\mciteBstWouldAddEndPuncttrue
\mciteSetBstMidEndSepPunct{\mcitedefaultmidpunct}
{\mcitedefaultendpunct}{\mcitedefaultseppunct}\relax
\EndOfBibitem
\bibitem[Ahmad-Bitar \emph{et~al.}(1977)Ahmad-Bitar, Lapatovich, Pritchard, and
  Renhorn]{bitar:NaNe}
R.~Ahmad-Bitar, W.~P. Lapatovich, D.~E. Pritchard and I.~Renhorn, \emph{Phys.
  Rev. Lett.}, 1977, \textbf{39}, 1657--1660\relax
\mciteBstWouldAddEndPuncttrue
\mciteSetBstMidEndSepPunct{\mcitedefaultmidpunct}
{\mcitedefaultendpunct}{\mcitedefaultseppunct}\relax
\EndOfBibitem
\bibitem[Brock and Duncan(1995)]{duncan:AgAr}
L.~R. Brock and M.~A. Duncan, \emph{J. Chem. Phys.}, 1995, \textbf{103},
  9200--9211\relax
\mciteBstWouldAddEndPuncttrue
\mciteSetBstMidEndSepPunct{\mcitedefaultmidpunct}
{\mcitedefaultendpunct}{\mcitedefaultseppunct}\relax
\EndOfBibitem
\bibitem[Jouvet \emph{et~al.}(1991)Jouvet, Lardeux-Dedonder, Martrenchard, and
  Solgadi]{jouvet:AgAr}
C.~Jouvet, C.~Lardeux-Dedonder, S.~Martrenchard and D.~Solgadi, \emph{J. Chem.
  Phys.}, 1991, \textbf{94}, 1759--1764\relax
\mciteBstWouldAddEndPuncttrue
\mciteSetBstMidEndSepPunct{\mcitedefaultmidpunct}
{\mcitedefaultendpunct}{\mcitedefaultseppunct}\relax
\EndOfBibitem
\bibitem[Brock and Duncan(1995)]{duncan:CuKr}
L.~R. Brock and M.~A. Duncan, \emph{Chem. Phys. Lett.}, 1995, \textbf{247},
  18--23\relax
\mciteBstWouldAddEndPuncttrue
\mciteSetBstMidEndSepPunct{\mcitedefaultmidpunct}
{\mcitedefaultendpunct}{\mcitedefaultseppunct}\relax
\EndOfBibitem
\bibitem[Cheng \emph{et~al.}(1989)Cheng, Willey, and Duncan]{duncan:Ag2Ar}
P.~Y. Cheng, K.~F. Willey and M.~A. Duncan, \emph{Chem. Phys. Lett.}, 1989,
  \textbf{163}, 469--474\relax
\mciteBstWouldAddEndPuncttrue
\mciteSetBstMidEndSepPunct{\mcitedefaultmidpunct}
{\mcitedefaultendpunct}{\mcitedefaultseppunct}\relax
\EndOfBibitem
\bibitem[Plowright \emph{et~al.}(2010)Plowright, Gardner, Withers, Wright,
  Morse, and Breckenridge]{breckenridge:AuNe}
R.~J. Plowright, A.~M. Gardner, C.~D. Withers, T.~G. Wright, M.~D. Morse and
  W.~H. Breckenridge, \emph{J. Phys. Chem. A}, 2010, \textbf{114}, 3103\relax
\mciteBstWouldAddEndPuncttrue
\mciteSetBstMidEndSepPunct{\mcitedefaultmidpunct}
{\mcitedefaultendpunct}{\mcitedefaultseppunct}\relax
\EndOfBibitem
\bibitem[Plowright \emph{et~al.}(2007)Plowright, Ayles, Watkins, Gardner,
  Wright, and Wright]{breckenridge:AuAr}
R.~J. Plowright, V.~L. Ayles, M.~J. Watkins, A.~M. Gardner, R.~R. Wright and
  T.~G. Wright, \emph{J. Chem. Phys.}, 2007, \textbf{127}, 204308\relax
\mciteBstWouldAddEndPuncttrue
\mciteSetBstMidEndSepPunct{\mcitedefaultmidpunct}
{\mcitedefaultendpunct}{\mcitedefaultseppunct}\relax
\EndOfBibitem
\bibitem[Bouchiat and Bouchiat(1970)]{bouchiat:alkalimolecules}
C.~C. Bouchiat and M.~A. Bouchiat, \emph{Phys. Rev. A}, 1970, \textbf{2},
  1274\relax
\mciteBstWouldAddEndPuncttrue
\mciteSetBstMidEndSepPunct{\mcitedefaultmidpunct}
{\mcitedefaultendpunct}{\mcitedefaultseppunct}\relax
\EndOfBibitem
\bibitem[Walker and Happer(1997)]{walker:rmp}
T.~G. Walker and W.~Happer, \emph{Rev. Mod. Phys.}, 1997, \textbf{69},
  629--642\relax
\mciteBstWouldAddEndPuncttrue
\mciteSetBstMidEndSepPunct{\mcitedefaultmidpunct}
{\mcitedefaultendpunct}{\mcitedefaultseppunct}\relax
\EndOfBibitem
\bibitem[Persson \emph{et~al.}(1996)Persson, Hui, Jakubek, Nakamura, and
  Takami]{takami:AgHe}
J.~L. Persson, Q.~Hui, Z.~J. Jakubek, M.~Nakamura and M.~Takami, \emph{Phys.
  Rev. Lett.}, 1996, \textbf{76}, 1501--1504\relax
\mciteBstWouldAddEndPuncttrue
\mciteSetBstMidEndSepPunct{\mcitedefaultmidpunct}
{\mcitedefaultendpunct}{\mcitedefaultseppunct}\relax
\EndOfBibitem
\bibitem[Takahashi \emph{et~al.}(1993)Takahashi, Sano, Kinoshita, and
  Yabuzaki]{takahashi:AlkHe1}
Y.~Takahashi, K.~Sano, T.~Kinoshita and T.~Yabuzaki, \emph{Phys. Rev. Lett.},
  1993, \textbf{71}, 1035--1038\relax
\mciteBstWouldAddEndPuncttrue
\mciteSetBstMidEndSepPunct{\mcitedefaultmidpunct}
{\mcitedefaultendpunct}{\mcitedefaultseppunct}\relax
\EndOfBibitem
\bibitem[Kojima \emph{et~al.}(1992)Kojima, Kobayashi, and Kaneko]{kojima:HHe+}
T.~M. Kojima, N.~Kobayashi and Y.~Kaneko, \emph{Z. Phys. D}, 1992, \textbf{23},
  181\relax
\mciteBstWouldAddEndPuncttrue
\mciteSetBstMidEndSepPunct{\mcitedefaultmidpunct}
{\mcitedefaultendpunct}{\mcitedefaultseppunct}\relax
\EndOfBibitem
\bibitem[Koyanagi and Bohme(2010)]{koyanagi:BaRg2+}
G.~K. Koyanagi and D.~K. Bohme, \emph{J. Phys. Chem. Lett.}, 2010, \textbf{1},
  41\relax
\mciteBstWouldAddEndPuncttrue
\mciteSetBstMidEndSepPunct{\mcitedefaultmidpunct}
{\mcitedefaultendpunct}{\mcitedefaultseppunct}\relax
\EndOfBibitem
\bibitem[Jakubek \emph{et~al.}(1997)Jakubek, Hui, and
  Takami]{takami:condensation}
Z.~J. Jakubek, Q.~Hui and M.~Takami, \emph{Phys. Rev. Lett.}, 1997,
  \textbf{79}, 629--632\relax
\mciteBstWouldAddEndPuncttrue
\mciteSetBstMidEndSepPunct{\mcitedefaultmidpunct}
{\mcitedefaultendpunct}{\mcitedefaultseppunct}\relax
\EndOfBibitem
\bibitem[Droppelmann \emph{et~al.}(2004)Droppelmann, B\"unermann, Schulz, and
  Stienkemeier]{droppelmann:RbHe}
G.~Droppelmann, O.~B\"unermann, C.~P. Schulz and F.~Stienkemeier, \emph{Phys.
  Rev. Lett.}, 2004, \textbf{93}, 023402\relax
\mciteBstWouldAddEndPuncttrue
\mciteSetBstMidEndSepPunct{\mcitedefaultmidpunct}
{\mcitedefaultendpunct}{\mcitedefaultseppunct}\relax
\EndOfBibitem
\bibitem[Mudrich \emph{et~al.}(2008)Mudrich, Stienkemeier, Droppelmann, Claas,
  and Schulz]{mudrich:RbHe}
M.~Mudrich, F.~Stienkemeier, G.~Droppelmann, P.~Claas and C.~P. Schulz,
  \emph{Phys. Rev. Lett.}, 2008, \textbf{100}, 023401\relax
\mciteBstWouldAddEndPuncttrue
\mciteSetBstMidEndSepPunct{\mcitedefaultmidpunct}
{\mcitedefaultendpunct}{\mcitedefaultseppunct}\relax
\EndOfBibitem
\bibitem[Reho \emph{et~al.}(2000)Reho, Higgins, Lehmann, and Scoles]{reho:KHe}
J.~Reho, J.~Higgins, K.~K. Lehmann and G.~Scoles, \emph{J. Chem. Phys.}, 2000,
  \textbf{113}, 9694\relax
\mciteBstWouldAddEndPuncttrue
\mciteSetBstMidEndSepPunct{\mcitedefaultmidpunct}
{\mcitedefaultendpunct}{\mcitedefaultseppunct}\relax
\EndOfBibitem
\bibitem[Reho \emph{et~al.}(1997)Reho, Callegari, Higgins, Ernst, Lehmann, and
  Scoles]{reho:NaHe}
J.~Reho, C.~Callegari, J.~Higgins, W.~E. Ernst, K.~K. Lehmann and G.~Scoles,
  \emph{Faraday Discuss.}, 1997, \textbf{108}, 161\relax
\mciteBstWouldAddEndPuncttrue
\mciteSetBstMidEndSepPunct{\mcitedefaultmidpunct}
{\mcitedefaultendpunct}{\mcitedefaultseppunct}\relax
\EndOfBibitem
\bibitem[Bartelt \emph{et~al.}(1996)Bartelt, Close, Federmann, Quaas, and
  Toennies]{toennies:AgHe}
A.~Bartelt, J.~D. Close, F.~Federmann, N.~Quaas and J.~P. Toennies, \emph{Phys.
  Rev. Lett.}, 1996, \textbf{77}, 3525--3528\relax
\mciteBstWouldAddEndPuncttrue
\mciteSetBstMidEndSepPunct{\mcitedefaultmidpunct}
{\mcitedefaultendpunct}{\mcitedefaultseppunct}\relax
\EndOfBibitem
\bibitem[Federmann \emph{et~al.}(1999)Federmann, Hoffmann, Quaas, and
  Close]{federmann:AgHe}
F.~Federmann, K.~Hoffmann, N.~Quaas and J.~D. Close, \emph{Phys. Rev. Lett.},
  1999, \textbf{83}, 2548--2551\relax
\mciteBstWouldAddEndPuncttrue
\mciteSetBstMidEndSepPunct{\mcitedefaultmidpunct}
{\mcitedefaultendpunct}{\mcitedefaultseppunct}\relax
\EndOfBibitem
\bibitem[Mella \emph{et~al.}(2002)Mella, Colombo, and Morosi]{mella:QMC}
M.~Mella, M.~C. Colombo and G.~Morosi, \emph{J. Chem. Phys.}, 2002,
  \textbf{117}, 9695\relax
\mciteBstWouldAddEndPuncttrue
\mciteSetBstMidEndSepPunct{\mcitedefaultmidpunct}
{\mcitedefaultendpunct}{\mcitedefaultseppunct}\relax
\EndOfBibitem
\bibitem[deCarvalho
  \emph{et~al.}(1999)deCarvalho\emph{et~al.}]{deCarvalho:bg:primer}
R.~deCarvalho \emph{et~al.}, \emph{Eur. Phys. J. D}, 1999, \textbf{7},
  289--309\relax
\mciteBstWouldAddEndPuncttrue
\mciteSetBstMidEndSepPunct{\mcitedefaultmidpunct}
{\mcitedefaultendpunct}{\mcitedefaultseppunct}\relax
\EndOfBibitem
\bibitem[Brahms \emph{et~al.}(2010)Brahms, Tscherbul, Zhang, K\l{}os,
  Sadeghpour, Dalgarno, Doyle, and Walker]{brahms:vdw}
N.~Brahms, T.~V. Tscherbul, P.~Zhang, J.~K\l{}os, H.~R. Sadeghpour,
  A.~Dalgarno, J.~M. Doyle and T.~G. Walker, \emph{Phys. Rev. Lett.}, 2010,
  \textbf{105}, 033001\relax
\mciteBstWouldAddEndPuncttrue
\mciteSetBstMidEndSepPunct{\mcitedefaultmidpunct}
{\mcitedefaultendpunct}{\mcitedefaultseppunct}\relax
\EndOfBibitem
\bibitem[Brahms \emph{et~al.}(2008)Brahms, Newman, Johnson, Greytak, Kleppner,
  and Doyle]{brahms:agTrapping}
N.~Brahms, B.~Newman, C.~Johnson, T.~Greytak, D.~Kleppner and J.~Doyle,
  \emph{Phys. Rev. Lett.}, 2008, \textbf{101}, 103002\relax
\mciteBstWouldAddEndPuncttrue
\mciteSetBstMidEndSepPunct{\mcitedefaultmidpunct}
{\mcitedefaultendpunct}{\mcitedefaultseppunct}\relax
\EndOfBibitem
\bibitem[Reif(1965)]{reif}
F.~Reif, \emph{Fundamentals of statistical and thermal physics}, McGraw-Hill,
  1965\relax
\mciteBstWouldAddEndPuncttrue
\mciteSetBstMidEndSepPunct{\mcitedefaultmidpunct}
{\mcitedefaultendpunct}{\mcitedefaultseppunct}\relax
\EndOfBibitem
\bibitem[Kleinekath{\"o}fer \emph{et~al.}(1999)Kleinekath{\"o}fer, Lewerenz,
  and Mladenovi{\'c}]{mladenovic:prl:alkaliBinding}
U.~Kleinekath{\"o}fer, M.~Lewerenz and M.~Mladenovi{\'c}, \emph{Phys. Rev.
  Lett.}, 1999, \textbf{83}, 4717\relax
\mciteBstWouldAddEndPuncttrue
\mciteSetBstMidEndSepPunct{\mcitedefaultmidpunct}
{\mcitedefaultendpunct}{\mcitedefaultseppunct}\relax
\EndOfBibitem
\bibitem[Brahms(2008)]{brahms:thesis}
N.~Brahms, \emph{PhD thesis}, Harvard Univ., 2008\relax
\mciteBstWouldAddEndPuncttrue
\mciteSetBstMidEndSepPunct{\mcitedefaultmidpunct}
{\mcitedefaultendpunct}{\mcitedefaultseppunct}\relax
\EndOfBibitem
\bibitem[Maxwell(2007)]{maxwell:thesis}
S.~Maxwell, \emph{PhD thesis}, Harvard Univ., 2007\relax
\mciteBstWouldAddEndPuncttrue
\mciteSetBstMidEndSepPunct{\mcitedefaultmidpunct}
{\mcitedefaultendpunct}{\mcitedefaultseppunct}\relax
\EndOfBibitem
\bibitem[Partridge \emph{et~al.}(2001)Partridge, Stallcop, and
  Levin]{dimerTable1}
H.~Partridge, J.~R. Stallcop and E.~Levin, \emph{J. Chem. Phys.}, 2001,
  \textbf{115}, 6471\relax
\mciteBstWouldAddEndPuncttrue
\mciteSetBstMidEndSepPunct{\mcitedefaultmidpunct}
{\mcitedefaultendpunct}{\mcitedefaultseppunct}\relax
\EndOfBibitem
\bibitem[Cargnoni \emph{et~al.}(2008)Cargnoni, Ku\'{s}, Mella, and
  Bartlett]{cargnoni:mHePotentials}
F.~Cargnoni, T.~Ku\'{s}, M.~Mella and R.~J. Bartlett, \emph{J. Chem. Phys.},
  2008, \textbf{129}, 204307\relax
\mciteBstWouldAddEndPuncttrue
\mciteSetBstMidEndSepPunct{\mcitedefaultmidpunct}
{\mcitedefaultendpunct}{\mcitedefaultseppunct}\relax
\EndOfBibitem
\bibitem[Krems \emph{et~al.}(2005)Krems, K\l{}os, Rode, Szcz\ifmmode
  \mbox{\c{e}}\else \c{e}\fi{}\ifmmode~\acute{s}\else \'{s}\fi{}niak,
  Cha\l{}asi\ifmmode~\acute{n}\else \'{n}\fi{}ski, and
  Dalgarno]{krems:prl:nons}
R.~V. Krems, J.~K\l{}os, M.~F. Rode, M.~M. Szcz\ifmmode \mbox{\c{e}}\else
  \c{e}\fi{}\ifmmode~\acute{s}\else \'{s}\fi{}niak,
  G.~Cha\l{}asi\ifmmode~\acute{n}\else \'{n}\fi{}ski and A.~Dalgarno,
  \emph{Phys. Rev. Lett.}, 2005, \textbf{94}, 013202\relax
\mciteBstWouldAddEndPuncttrue
\mciteSetBstMidEndSepPunct{\mcitedefaultmidpunct}
{\mcitedefaultendpunct}{\mcitedefaultseppunct}\relax
\EndOfBibitem
\bibitem[Maxwell \emph{et~al.}(2008)Maxwell, Hummon, Wang, Buchachenko, Krems,
  and Doyle]{maxwell:Bi:pra08}
S.~E. Maxwell, M.~T. Hummon, Y.~Wang, A.~A. Buchachenko, R.~V. Krems and J.~M.
  Doyle, \emph{Phys. Rev. A}, 2008, \textbf{78}, 042706\relax
\mciteBstWouldAddEndPuncttrue
\mciteSetBstMidEndSepPunct{\mcitedefaultmidpunct}
{\mcitedefaultendpunct}{\mcitedefaultseppunct}\relax
\EndOfBibitem
\bibitem[Cybulski \emph{et~al.}(2005)Cybulski\emph{et~al.}]{Cybulski2005:NHHe}
H.~Cybulski \emph{et~al.}, \emph{J. Chem. Phys.}, 2005, \textbf{122},
  094307\relax
\mciteBstWouldAddEndPuncttrue
\mciteSetBstMidEndSepPunct{\mcitedefaultmidpunct}
{\mcitedefaultendpunct}{\mcitedefaultseppunct}\relax
\EndOfBibitem
\bibitem[Groenenboom and Balakrishnan(2003)]{Groenenboom2003:CaHHe}
G.~C. Groenenboom and N.~Balakrishnan, \emph{J. Chem. Phys.}, 2003,
  \textbf{118}, 7380--7385\relax
\mciteBstWouldAddEndPuncttrue
\mciteSetBstMidEndSepPunct{\mcitedefaultmidpunct}
{\mcitedefaultendpunct}{\mcitedefaultseppunct}\relax
\EndOfBibitem
\bibitem[Tscherbul \emph{et~al.}(2007)Tscherbul, K{\l}os, Rajchel, and
  Krems]{krems:ybf:pra2007}
T.~V. Tscherbul, J.~K{\l}os, L.~Rajchel and R.~V. Krems, \emph{Phys. Rev. A},
  2007, \textbf{75}, 033416\relax
\mciteBstWouldAddEndPuncttrue
\mciteSetBstMidEndSepPunct{\mcitedefaultmidpunct}
{\mcitedefaultendpunct}{\mcitedefaultseppunct}\relax
\EndOfBibitem
\bibitem[Turpin \emph{et~al.}(2010)Turpin, Halvick, and Stoecklin]{turpin:MnH}
F.~Turpin, P.~Halvick and T.~Stoecklin, \emph{J. Chem. Phys.}, 2010,
  \textbf{132}, 214305\relax
\mciteBstWouldAddEndPuncttrue
\mciteSetBstMidEndSepPunct{\mcitedefaultmidpunct}
{\mcitedefaultendpunct}{\mcitedefaultseppunct}\relax
\EndOfBibitem
\bibitem[Jakubek and Takami(1997)]{jakubek:agHe}
Z.~J. Jakubek and M.~Takami, \emph{Chem. Phys. Lett.}, 1997, \textbf{265}, 653
  -- 659\relax
\mciteBstWouldAddEndPuncttrue
\mciteSetBstMidEndSepPunct{\mcitedefaultmidpunct}
{\mcitedefaultendpunct}{\mcitedefaultseppunct}\relax
\EndOfBibitem
\bibitem[Cargnoni and Mella(2011)]{cargnoni:2011}
F.~Cargnoni and M.~Mella, \emph{J. Phys. Chem. A}, 2011,  in press\relax
\mciteBstWouldAddEndPuncttrue
\mciteSetBstMidEndSepPunct{\mcitedefaultmidpunct}
{\mcitedefaultendpunct}{\mcitedefaultseppunct}\relax
\EndOfBibitem
\bibitem[Knowles \emph{et~al.}(1993)Knowles, Hampel, and Werner]{knowles:combo}
P.~J. Knowles, C.~Hampel and H.-J. Werner, \emph{J. Chem. Phys.}, 1993,
  \textbf{99}, 5219\relax
\mciteBstWouldAddEndPuncttrue
\mciteSetBstMidEndSepPunct{\mcitedefaultmidpunct}
{\mcitedefaultendpunct}{\mcitedefaultseppunct}\relax
\EndOfBibitem
\bibitem[Werner \emph{et~al.}(2008)Werner\emph{et~al.}]{MOLPRO}
H.-J. Werner \emph{et~al.}, \emph{{\textsc{molpro}}}, 2008, {URL
  \texttt{http://www.molpro.net}}\relax
\mciteBstWouldAddEndPuncttrue
\mciteSetBstMidEndSepPunct{\mcitedefaultmidpunct}
{\mcitedefaultendpunct}{\mcitedefaultseppunct}\relax
\EndOfBibitem
\bibitem[Woon and T.H.~Dunning(1994)]{woon:94}
D.~Woon and J.~T.H.~Dunning, \emph{J. Chem. Phys.}, 1994, \textbf{100},
  2975\relax
\mciteBstWouldAddEndPuncttrue
\mciteSetBstMidEndSepPunct{\mcitedefaultmidpunct}
{\mcitedefaultendpunct}{\mcitedefaultseppunct}\relax
\EndOfBibitem
\bibitem[Figgen \emph{et~al.}(2005)Figgen, Rauhut, Dolg, and Stoll]{figgen:05}
D.~Figgen, G.~Rauhut, M.~Dolg and H.~Stoll, \emph{Chem. Phys.}, 2005,
  \textbf{311}, 227 -- 244\relax
\mciteBstWouldAddEndPuncttrue
\mciteSetBstMidEndSepPunct{\mcitedefaultmidpunct}
{\mcitedefaultendpunct}{\mcitedefaultseppunct}\relax
\EndOfBibitem
\bibitem[Peterson and Puzzarini(2005)]{peterson:05}
K.~Peterson and C.~Puzzarini, \emph{Theor. Chem. Acc}, 2005, \textbf{114},
  283\relax
\mciteBstWouldAddEndPuncttrue
\mciteSetBstMidEndSepPunct{\mcitedefaultmidpunct}
{\mcitedefaultendpunct}{\mcitedefaultseppunct}\relax
\EndOfBibitem
\bibitem[Boys and Bernardi(1970)]{boys:70}
S.~F. Boys and F.~Bernardi, \emph{Mol. Phys.}, 1970, \textbf{19},
  553--566\relax
\mciteBstWouldAddEndPuncttrue
\mciteSetBstMidEndSepPunct{\mcitedefaultmidpunct}
{\mcitedefaultendpunct}{\mcitedefaultseppunct}\relax
\EndOfBibitem
\bibitem[Tong \emph{et~al.}(2009)Tong, Yang, An, Wang, Ma, and Wang]{tong:09}
X.-F. Tong, C.-L. Yang, Y.-P. An, M.-S. Wang, X.-G. Ma and D.-H. Wang, \emph{J.
  Chem. Phys.}, 2009, \textbf{131}, 244304\relax
\mciteBstWouldAddEndPuncttrue
\mciteSetBstMidEndSepPunct{\mcitedefaultmidpunct}
{\mcitedefaultendpunct}{\mcitedefaultseppunct}\relax
\EndOfBibitem
\bibitem[Gardner \emph{et~al.}(2010)Gardner, Plowright, Watkins, Wright, and
  Breckenridge]{gardner:agHe:10}
A.~M. Gardner, R.~J. Plowright, M.~J. Watkins, T.~G. Wright and W.~H.
  Breckenridge, \emph{J. Chem. Phys.}, 2010, \textbf{132}, 184301\relax
\mciteBstWouldAddEndPuncttrue
\mciteSetBstMidEndSepPunct{\mcitedefaultmidpunct}
{\mcitedefaultendpunct}{\mcitedefaultseppunct}\relax
\EndOfBibitem
\bibitem[Colbert and Miller(1992)]{colbert:dvr:92}
D.~T. Colbert and W.~H. Miller, \emph{J. Chem. Phys.}, 1992, \textbf{96},
  1982\relax
\mciteBstWouldAddEndPuncttrue
\mciteSetBstMidEndSepPunct{\mcitedefaultmidpunct}
{\mcitedefaultendpunct}{\mcitedefaultseppunct}\relax
\EndOfBibitem
\bibitem[Carrington \emph{et~al.}(1995)Carrington, Leach, Marr, Shaw, Viant,
  Hutson, and Law]{carrington:jcp:94}
A.~Carrington, C.~A. Leach, A.~J. Marr, A.~M. Shaw, M.~R. Viant, J.~M. Hutson
  and M.~M. Law, \emph{J. Chem. Phys.}, 1995, \textbf{102}, 2379\relax
\mciteBstWouldAddEndPuncttrue
\mciteSetBstMidEndSepPunct{\mcitedefaultmidpunct}
{\mcitedefaultendpunct}{\mcitedefaultseppunct}\relax
\EndOfBibitem
\bibitem[Buchachenko \emph{et~al.}(2003)Buchachenko, Grinev, K{\l}os, Bieske,
  Szcz\c{e}{\'s}niak, , and Cha{\l}asi{\'n}ski]{grinev:jcp03}
A.~A. Buchachenko, T.~A. Grinev, J.~K{\l}os, E.~J. Bieske, M.~M.
  Szcz\c{e}{\'s}niak,  and G.~Cha{\l}asi{\'n}ski, \emph{J. Chem. Phys.}, 2003,
  \textbf{119}, 12931\relax
\mciteBstWouldAddEndPuncttrue
\mciteSetBstMidEndSepPunct{\mcitedefaultmidpunct}
{\mcitedefaultendpunct}{\mcitedefaultseppunct}\relax
\EndOfBibitem
\bibitem[Grimme(2006)]{hyperfine:Abinitio1}
S.~Grimme, \emph{J. Chem. Phys.}, 2006, \textbf{124}, 034108\relax
\mciteBstWouldAddEndPuncttrue
\mciteSetBstMidEndSepPunct{\mcitedefaultmidpunct}
{\mcitedefaultendpunct}{\mcitedefaultseppunct}\relax
\EndOfBibitem
\bibitem[van Lenthe \emph{et~al.}(1993)van Lenthe, Baerends, and
  Snijders]{hyperfine:Abinitio2}
E.~van Lenthe, E.~J. Baerends and J.~G. Snijders, \emph{J. Chem. Phys.}, 1993,
  \textbf{99}, 4597\relax
\mciteBstWouldAddEndPuncttrue
\mciteSetBstMidEndSepPunct{\mcitedefaultmidpunct}
{\mcitedefaultendpunct}{\mcitedefaultseppunct}\relax
\EndOfBibitem
\bibitem[van W{\"u}llen(1998)]{hyperfine:Abinitio3}
C.~van W{\"u}llen, \emph{J. Chem. Phys.}, 1998, \textbf{109}, 392\relax
\mciteBstWouldAddEndPuncttrue
\mciteSetBstMidEndSepPunct{\mcitedefaultmidpunct}
{\mcitedefaultendpunct}{\mcitedefaultseppunct}\relax
\EndOfBibitem
\bibitem[Belanzoni \emph{et~al.}(1998)Belanzoni, van Lenthe, and
  Baerends]{hyperfine:Abinitio4}
P.~Belanzoni, E.~van Lenthe and E.~J. Baerends, \emph{J. Chem. Phys.}, 1998,
  \textbf{114}, 4421--4424\relax
\mciteBstWouldAddEndPuncttrue
\mciteSetBstMidEndSepPunct{\mcitedefaultmidpunct}
{\mcitedefaultendpunct}{\mcitedefaultseppunct}\relax
\EndOfBibitem
\bibitem[van Lenthe \emph{et~al.}(1998)van Lenthe, van~der Avoird, and
  Wormer]{hyperfine:Abinitio5}
E.~van Lenthe, A.~van~der Avoird and P.~E.~S. Wormer, \emph{J. Chem. Phys.},
  1998, \textbf{108}, 4783--4796\relax
\mciteBstWouldAddEndPuncttrue
\mciteSetBstMidEndSepPunct{\mcitedefaultmidpunct}
{\mcitedefaultendpunct}{\mcitedefaultseppunct}\relax
\EndOfBibitem
\bibitem[Huzinaga and Klobukowski(1993)]{hyperfine:Abinitio6}
S.~Huzinaga and M.~Klobukowski, \emph{Chem. Phys. Lett.}, 1993, \textbf{212},
  260--264\relax
\mciteBstWouldAddEndPuncttrue
\mciteSetBstMidEndSepPunct{\mcitedefaultmidpunct}
{\mcitedefaultendpunct}{\mcitedefaultseppunct}\relax
\EndOfBibitem
\bibitem[Jensen(2007)]{hyperfine:Abinitio7}
F.~Jensen, \emph{J. Phys. Chem. A}, 2007, \textbf{111}, 11198\relax
\mciteBstWouldAddEndPuncttrue
\mciteSetBstMidEndSepPunct{\mcitedefaultmidpunct}
{\mcitedefaultendpunct}{\mcitedefaultseppunct}\relax
\EndOfBibitem
\bibitem[Ting and Lew(1957)]{hyperfine:Abinitio9}
Y.~Ting and H.~Lew, \emph{Phys. Rev.}, 1957, \textbf{105}, 581\relax
\mciteBstWouldAddEndPuncttrue
\mciteSetBstMidEndSepPunct{\mcitedefaultmidpunct}
{\mcitedefaultendpunct}{\mcitedefaultseppunct}\relax
\EndOfBibitem
\bibitem[Walker \emph{et~al.}(1997)Walker, Thywissen, and
  Happer]{walker:pra1997:spinRotation}
T.~G. Walker, J.~H. Thywissen and W.~Happer, \emph{Phys. Rev. A}, 1997,
  \textbf{56}, 2090\relax
\mciteBstWouldAddEndPuncttrue
\mciteSetBstMidEndSepPunct{\mcitedefaultmidpunct}
{\mcitedefaultendpunct}{\mcitedefaultseppunct}\relax
\EndOfBibitem
\bibitem[Saha(1993)]{saha:scattering_length}
H.~P. Saha, \emph{Phys. Rev. A}, 1993, \textbf{48}, 1163\relax
\mciteBstWouldAddEndPuncttrue
\mciteSetBstMidEndSepPunct{\mcitedefaultmidpunct}
{\mcitedefaultendpunct}{\mcitedefaultseppunct}\relax
\EndOfBibitem
\bibitem[Clementi and Roetti(1974)]{roetti:rhf}
E.~Clementi and C.~Roetti, \emph{Atom. Data Nucl. Data Tables}, 1974,
  \textbf{14}, 177--478\relax
\mciteBstWouldAddEndPuncttrue
\mciteSetBstMidEndSepPunct{\mcitedefaultmidpunct}
{\mcitedefaultendpunct}{\mcitedefaultseppunct}\relax
\EndOfBibitem
\bibitem[Tscherbul \emph{et~al.}(2008)Tscherbul, Zhang, Sadeghpour, Dalgarno,
  Brahms, Au, and Doyle]{tscherbul:AlkHe:2008}
T.~V. Tscherbul, P.~Zhang, H.~Sadeghpour, A.~Dalgarno, N.~Brahms, Y.~S. Au and
  J.~M. Doyle, \emph{Phys. Rev. A}, 2008, \textbf{78}, 060703(R)\relax
\mciteBstWouldAddEndPuncttrue
\mciteSetBstMidEndSepPunct{\mcitedefaultmidpunct}
{\mcitedefaultendpunct}{\mcitedefaultseppunct}\relax
\EndOfBibitem
\bibitem[Tscherbul \emph{et~al.}(2009)Tscherbul, Zhang, Sadeghpour, and
  Dalgarno]{tscherbul:AlkHe:2009}
T.~V. Tscherbul, P.~Zhang, H.~R. Sadeghpour and A.~Dalgarno, \emph{Phys. Rev.
  A}, 2009, \textbf{79}, 062707\relax
\mciteBstWouldAddEndPuncttrue
\mciteSetBstMidEndSepPunct{\mcitedefaultmidpunct}
{\mcitedefaultendpunct}{\mcitedefaultseppunct}\relax
\EndOfBibitem
\bibitem[Roberts \emph{et~al.}(1969)Roberts, Bernstein, and
  Curtiss]{rbc:jcp:recombination}
R.~E. Roberts, R.~B. Bernstein and C.~F. Curtiss, \emph{J. Chem. Phys.}, 1969,
  \textbf{50}, 5163--5176\relax
\mciteBstWouldAddEndPuncttrue
\mciteSetBstMidEndSepPunct{\mcitedefaultmidpunct}
{\mcitedefaultendpunct}{\mcitedefaultseppunct}\relax
\EndOfBibitem
\bibitem[Paolini \emph{et~al.}(2011)Paolini, Ohlinger, and
  Forrey]{paolini:2011}
S.~Paolini, L.~Ohlinger and R.~C. Forrey, \emph{Phys. Rev. A}, 2011,  in
  press\relax
\mciteBstWouldAddEndPuncttrue
\mciteSetBstMidEndSepPunct{\mcitedefaultmidpunct}
{\mcitedefaultendpunct}{\mcitedefaultseppunct}\relax
\EndOfBibitem
\bibitem[Hutson and Green(1994)]{molscat}
J.~M. Hutson and S.~Green, 1994, \texttt{MOLSCAT} computer code, version 14,
  distributed by Collaborative Computational Project No. 6 of the Engineering
  and Physical Sciences Research Council (UK)\relax
\mciteBstWouldAddEndPuncttrue
\mciteSetBstMidEndSepPunct{\mcitedefaultmidpunct}
{\mcitedefaultendpunct}{\mcitedefaultseppunct}\relax
\EndOfBibitem
\bibitem[Jeziorska \emph{et~al.}(2007)Jeziorska, Cencek, Patkowski, Jeziorski,
  and Szalewicz]{jeziorska:hedimer}
M.~Jeziorska, W.~Cencek, K.~Patkowski, B.~Jeziorski and K.~Szalewicz, \emph{J.
  Chem. Phys.}, 2007, \textbf{127}, 124303\relax
\mciteBstWouldAddEndPuncttrue
\mciteSetBstMidEndSepPunct{\mcitedefaultmidpunct}
{\mcitedefaultendpunct}{\mcitedefaultseppunct}\relax
\EndOfBibitem
\bibitem[Miller(1986)]{miller:vdw:review}
R.~E. Miller, \emph{J. Chem. Phys.}, 1986, \textbf{90}, 3301--3313\relax
\mciteBstWouldAddEndPuncttrue
\mciteSetBstMidEndSepPunct{\mcitedefaultmidpunct}
{\mcitedefaultendpunct}{\mcitedefaultseppunct}\relax
\EndOfBibitem
\bibitem[Demtr{\"o}der(2002)]{demtroder}
W.~Demtr{\"o}der, \emph{Laser Spectroscopy}, Springer-Verlag, 3rd edn,
  2002\relax
\mciteBstWouldAddEndPuncttrue
\mciteSetBstMidEndSepPunct{\mcitedefaultmidpunct}
{\mcitedefaultendpunct}{\mcitedefaultseppunct}\relax
\EndOfBibitem
\bibitem[Br\"uhl \emph{et~al.}(2005)Br\"uhl, Kalinin, Kornilov, Toennies,
  Hegerfeldt, and Stoll]{stoll:he3}
R.~Br\"uhl, A.~Kalinin, O.~Kornilov, J.~P. Toennies, G.~C. Hegerfeldt and
  M.~Stoll, \emph{Phys. Rev. Lett.}, 2005, \textbf{95}, 063002\relax
\mciteBstWouldAddEndPuncttrue
\mciteSetBstMidEndSepPunct{\mcitedefaultmidpunct}
{\mcitedefaultendpunct}{\mcitedefaultseppunct}\relax
\EndOfBibitem
\bibitem[Gonz{\'a}lez-Lezana \emph{et~al.}(1999)Gonz{\'a}lez-Lezana,
  Rubayo-Soneira, Miret-Art{\'e}s, Gianturco, Delgado-Barrio, and
  Villareal]{villareal:he3}
T.~Gonz{\'a}lez-Lezana, J.~Rubayo-Soneira, S.~Miret-Art{\'e}s, F.~A. Gianturco,
  G.~Delgado-Barrio and P.~Villareal, \emph{Phys. Rev. Lett.}, 1999,
  \textbf{82}, 1648\relax
\mciteBstWouldAddEndPuncttrue
\mciteSetBstMidEndSepPunct{\mcitedefaultmidpunct}
{\mcitedefaultendpunct}{\mcitedefaultseppunct}\relax
\EndOfBibitem
\bibitem[Esry \emph{et~al.}(2001)Esry, Lin, Greene, and Blume]{esry:He3}
B.~D. Esry, C.~D. Lin, C.~H. Greene and D.~Blume, \emph{Phys. Rev. Lett.},
  2001, \textbf{86}, 4189\relax
\mciteBstWouldAddEndPuncttrue
\mciteSetBstMidEndSepPunct{\mcitedefaultmidpunct}
{\mcitedefaultendpunct}{\mcitedefaultseppunct}\relax
\EndOfBibitem
\end{mcitethebibliography}
\end{document}